\documentclass[journal]{IEEEtran}
\usepackage{amsmath,amsfonts}
\usepackage[linesnumbered,ruled,vlined]{algorithm2e}
\usepackage{booktabs}
\usepackage{graphicx}
\usepackage{multirow}
\usepackage{array}

\newcommand{\vect}{\mathbf}
\newcommand{\HT}{\mathsf{H}}

\newcommand{\OL}{\overline}
\newcommand{\LOL}[1]{\underline{{#1}}}
\newcommand{\DNDRSS}{DN+DR+SS}
\newcommand{\CBF}{CBF}
\newcommand{\BF}{beamformer}
\newcommand{\form}{implementation}
\newcommand{\MIMO}{source-packed}
\newcommand{\Mimo}{Source-packed}
\newcommand{\SW}{source-wise}
\newcommand{\Sw}{Source-wise}
\newcommand{\tw}{MISO}
\newcommand{\TW}{MISO}
\newcommand{\target}{single-target}
\usepackage{color}

\newcommand{\mcc}[1]{\multicolumn{1}{c}{#1}}

\usepackage[absolute,showboxes]{textpos}
\TPMargin{5pt}
\newcommand{\copyrightstatement}{
\begin{textblock}{0.8}(0.1,0.01)
\noindent
\footnotesize
\copyright 2020 IEEE.  Personal use of this material is permitted.  Permission from IEEE must be obtained for all other uses, in any current or future media, including reprinting/republishing this material for advertising or promotional purposes, creating new collective works, for resale or redistribution to servers or lists, or reuse of any copyrighted component of this work in other works.
\end{textblock}
}
\begin{document}
\copyrightstatement

\title{Jointly optimal denoising, dereverberation, \\and source separation}
\author{Tomohiro Nakatani,~\IEEEmembership{Senior Member,~IEEE,}
        Christoph Boeddeker,~\IEEEmembership{Student Member,~IEEE,}\\
        Keisuke Kinoshita,~\IEEEmembership{Senior Member,~IEEE,}
        Rintaro Ikeshita,~\IEEEmembership{Member,~IEEE,}\\
        Marc Delcroix,~\IEEEmembership{Senior Member,~IEEE,}
        Reinhold Haeb-Umbach,~\IEEEmembership{Fellow,~IEEE}
\thanks{T. Nakatani, K. Kinoshita, R. Ikeshita, and M. Delcroix are with NTT Corporation. C. Boeddeker and R. Haeb-Umbach are with Paderborn Univ.}
\thanks{Manuscript received January 1, 2020; revised XXXX XX, 2020.}}




\maketitle

\begin{abstract}
This paper proposes methods that can optimize a Convolutional BeamFormer ({\CBF}) for {jointly} performing denoising, dereverberation, and source separation ({\DNDRSS}) {in a computationally efficient way}. 
%
Conventionally, a cascade configuration, composed of a Weighted Prediction Error minimization (WPE) dereverberation filter followed by a Minimum Variance Distortionless Response (MVDR) beamformer, has been used as the state-of-the-art frontend of far-field speech recognition,
even though this approach's overall optimality is not guaranteed.  
In the blind signal processing area, an approach for jointly optimizing dereverberation and source separation (DR+SS) has been proposed; however, it requires huge computing cost, and has not been extended for applications to {\DNDRSS}. 
%
To overcome the above limitations, this paper develops new approaches for {jointly} optimizing {\DNDRSS} in a computationally much more efficient way. 
{To this end, we first present an objective function to optimize a {\CBF} for performing DN+DR+SS based on maximum likelihood estimation on an assumption that the steering vectors of the target signals are given or can be estimated, e.g., using a neural network. This paper refers to a {\CBF} optimized by this objective function as a weighted Minimum-Power Distortionless Response (wMPDR) {\CBF}. Then, we derive two algorithms for optimizing a wMPDR {\CBF} based on two different ways of factorizing a {\CBF} into WPE filters and {\BF}s: one based on an extension of the conventional joint optimization approach proposed for DR+SS and another based on a novel technique.}
%
%
%
Experiments using noisy reverberant sound mixtures 
show that the proposed optimization approaches greatly improve the performance of the speech enhancement in comparison with the conventional cascade configuration in terms of signal distortion measures and ASR performance. The proposed approaches also greatly reduce the computing cost with improved estimation accuracy in comparison with the conventional joint optimization approach. 
\end{abstract}

\IEEEpeerreviewmaketitle

\section{Introduction}
When a speech signal is captured by distant microphones, e.g., in a conference room, it often contains reverberation, diffuse noise, and extraneous speakers' voices. These components are detrimental to the intelligibility of the captured speech and often cause serious degradation in many applications such as hands-free teleconferencing and Automatic Speech Recognition (ASR).

Microphone array speech enhancement has been scrutinized to minimize the aforementioned detrimental effects in the acquired signal. For performing denoising (DN), beamforming techniques have been investigated for decades \cite{Veen88ASSP,MPDR,Cox,Souden07ASLP}, and the Minimum Variance Distortionless Response (MVDR) beamformer and the Minimum Power Distortionless Response (MPDR) {\BF}, are now widely used as state-of-the-art techniques. For source separation (SS), a number of blind signal processing techniques have been developed, including independent component analysis \cite{ica}, independent vector analysis \cite{iva}, and spatial clustering-based beamforming \cite{souden10aslp}. For dereverberation (DR), a Weighted Prediction Error minimization (WPE) based linear prediction technique \cite{wpe,gwpe} and its variants \cite{swpe} have been actively studied as an effective approach. With these techniques, for determining the coefficients of filtering, it is crucial to accurately estimate such statistics of the speech signals and the noise as their spatial covariances and time-varying variances. However, the estimation often becomes inaccurate when the signals are mixed under reverberant and noisy conditions, which seriously degrades the performance of these techniques.

To enhance the robustness of the above techniques, neural network-supported microphone array speech enhancement has been actively studied, and its effectiveness has been identified for denoising \cite{heymann16icassp}, dereverberation \cite{DNN-WPE}, and source separation \cite{speakerbeam,yoshioka2018interspeech}.  With this approach, neural networks estimate such statistics of the signals and noise as Time-Frequency (TF) masks and time-varying variances \cite{speakerbeam,YongXu,dc16icassp,pit17taslp}, while microphone array signal processing performs speech enhancement. This combination is particularly effective because neural networks can successfully capture the spectral patterns of signals over wide TF ranges and reliably estimate such statistics of the signals.  Conventional signal processing often fails to adequately handle them. On the other hand, neural networks often introduce into the processed signal nonlinear distortions, which are harmful to perceived speech quality and ASR. This problem can be avoided by microphone array techniques. A number of articles have reported the usefulness of this combination, particularly for far-field ASR, e.g., at the REVERB challenge \cite{REVERB} and the CHiME-3/4/5 challenges \cite{chime3,Kanda2019}.

Despite the success of neural network-supported microphone array speech enhancement, how to optimally combine individual microphone array techniques for simultaneously performing denoising, dereverberation, and source separation ({\DNDRSS}) in a computationally efficient way remains inadequately investigated. 
For example, for denoising and dereverberation (DN+DR), the cascade configuration of a WPE filter followed by a MVDR/MPDR {\BF} has been widely used as the state-of-the-art frontend, e.g., at the far-field ASR challenges \cite{REVERB,chime3,Kanda2019,SPM}. However, since the WPE filter and the {\BF} are separately optimized, the overall optimality of this approach is not guaranteed.
To optimally perform DN+DR, several techniques have been proposed using a Kalman filter \cite{togami,Braun2018taslp,wpegsc}. 
A technique, called Integrated Sidelobe Cancellation and Linear Prediction (ISCLP) \cite{wpegsc}, optimizes an integrated filter that can cancel noise and reverberation from the observed signals using a sidelobe cancellation framework.  With this technique, however, the steering vector of the target signal needs to be directly estimated in advance from noisy reverberant speech, which is challenging and limits the overall estimation accuracy.
In the blind signal processing area, on the other hand, a technique that jointly optimizes a pair comprised of a WPE filter and a beamformer has been proposed for dereverberation and source separation (DR+SS) under noiseless conditions \cite{takuya2011taslp,ito2014iwaenc,Kagami2018icassp}. 
One advantage of this approach is that we can access multichannel dereverberated signals obtained as the output of the WPE filter during the optimization, and utilize them to reliably estimate the beamformer. However, this approach requires 1) huge computing cost for the optimization, and 2) has not been extended for application to {\DNDRSS}. 

{To overcome the above limitations, this paper develops algorithms for optimizing a Convolutional BeamFormer (\CBF) that can perform {\DNDRSS} in a computationally much more efficient way. 
A {\CBF} is a filter that is applied to a multichannel observed signal to yield the desired output signals. 
For {\CBF} optimization, this paper first presents a common objective function based on the Maximum Likelihood (ML) criterion by assuming that the steering vectors of the desired signals are given, or can be estimated. This paper refers to a {\CBF} optimized by this objective function as a weighted MPDR (wMPDR) {\CBF}. After showing that a {\CBF} can be factorized into WPE filter(s) and beamformer(s) in two different ways, we derive two different algorithms for optimizing the wMPDR {\CBF}, based on the {\CBF} factorization ways.}
%
The first approach, called {\MIMO} factorization, is an extension of the conventional joint optimization technique proposed for DR+SS \cite{takuya2011taslp,ito2014iwaenc,Kagami2018icassp}. We first show that its direct application to {\DNDRSS} suffers from serious problems in terms of the computational efficiency and estimation accuracy and present an extension for solving them. 
The second approach, called {\SW} factorization, is based on a novel factorization technique that factorizes a {\CBF} into a set of sub-filter pairs, each of which is composed of a WPE filter and a beamformer, and independently estimates each source.
{For both approaches, we also present a method that robustly estimates the steering vectors of the desired signals during the wMPDR {\CBF} optimization using the output of the WPE filters.} A neural network-supported TF-mask estimation technique is also incorporated\footnote{Note that the proposed techniques can also be applied to conventional blind signal processing for DR+SS, as discussed in an article {\cite{nak2020interspeech}}.} to estimate the steering vectors. 
%
%
Although both approaches work comparably well in terms of estimation accuracy, {\SW} factorization has advantages in terms of computational efficiency. 
An additional benefit of {\SW} factorization is that it can be used, without loss of optimality for the extraction of a single target source from a sound mixture, which is now an important application area of speech enhancement \cite{speakerbeam,koldovsky2019tsp}. 

Experiments based on noisy reverberant sound mixtures created using the REVERB Challenge dataset \cite{REVERB} show that the proposed optimization approaches substantially improve the {\DNDRSS} performance in comparison to the conventional cascade configuration in terms of ASR performance and signal distortion reduction. These two proposed approaches can also greatly reduce the computing cost with improved estimation accuracy in comparison with the conventional joint optimization approach.

Certain parts of this paper have already been presented in our recent conference papers.  The ML formulation for optimizing a {\CBF} was derived for DN+DR \cite{MLWPD}. Another work \cite{factorizedwpd} argued that a {\CBF} for DN+DR can be factorized into a WPE filter and a wMPDR (non-convolutional) beamformer, and jointly optimized without loss of optimality. Another work \cite{nak2020icassp} presented ways to reliably estimate TF masks for {\DNDRSS}. This paper integrates these techniques to perform {\DNDRSS} {in a computationally efficient way}.

In the remainder of this paper, the models of the observed signal and the {\CBF} are defined in Section~\ref{sec:model}. Then, Section~\ref{sec:mlcb} presents our proposed optimization methods, and  Section~\ref{sec:discussion} summarizes their characteristics and advantages. Sections~\ref{sec:exp} and \ref{sec:conclusion} describe experimental results and concluding remarks.

\section{Models of signal and beamformer}\label{sec:model}
\begin{figure*}[t]
\centering
\begin{tabular}{ccc}
 {\includegraphics[width=5.6cm]{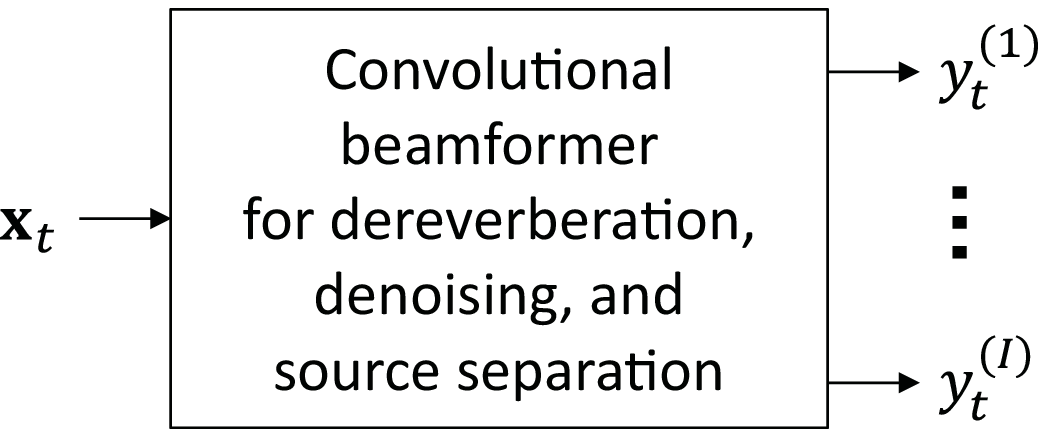}} &&
{\includegraphics[width=6.6cm]{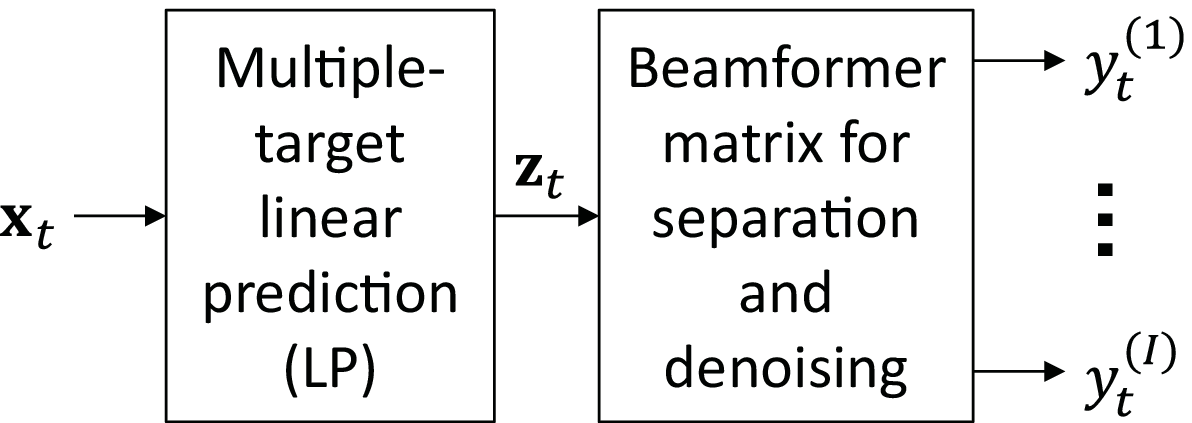}} \\
(a) {MIMO \CBF} &&(b) MIMO CBF with {\MIMO} factorization \\\smallskip\\
{\includegraphics[width=5.76cm]{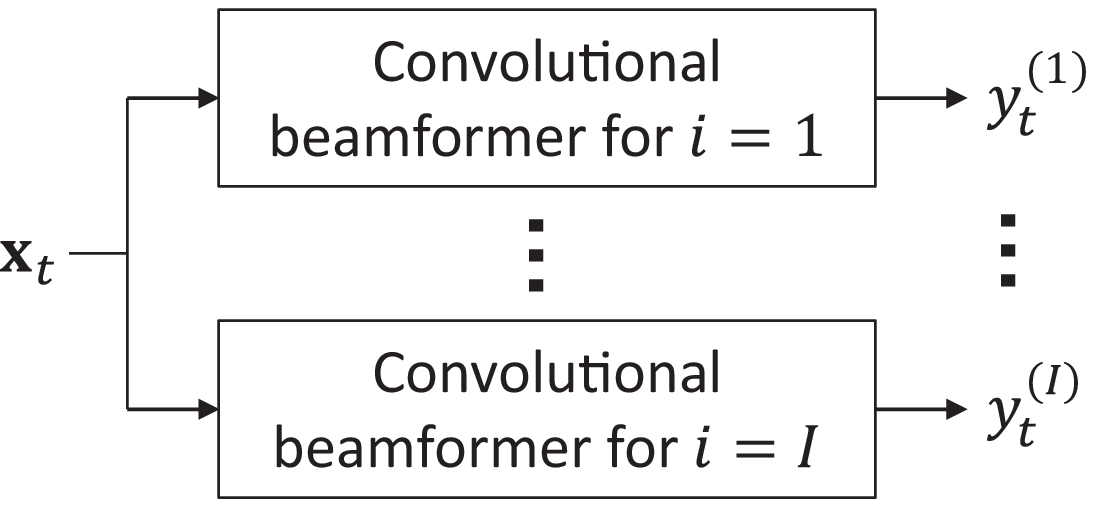}} &&\includegraphics[width=6.84cm]{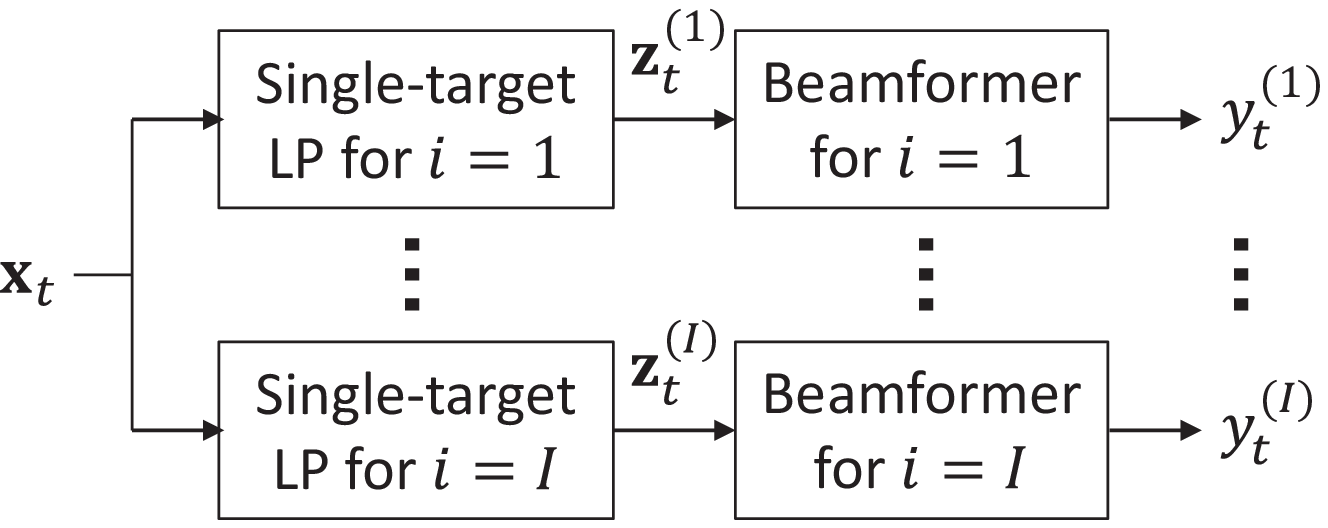}\vspace{-6mm}\\
(c) {Set of {\TW} {\CBF}s} && (d) MISO CBFs with {\SW} factorization\\
\end{tabular}
\caption{Multi-Input Multi-Output (MIMO) {\CBF} and its three different {\form}s. They are equivalent to each other in the sense that whatever values are set to coefficients of one {\form}, certain coefficients of the other {\form}s can be determined such that they realize identical input-output relationships. Thus, optimal solutions of all {\form}s are identical as long as they are optimized based on the same objective function.}\label{fig:imples}
\end{figure*}

This paper assumes that $I$ source signals are captured by $M(\ge I)$ microphones in a noisy reverberant environment. The captured signal at each TF point in the short-time Fourier transform (STFT) domain is modeled by 
\begin{align}
\vect{x}_{t,f}&=\sum_{i=1}^I\vect{x}_{t,f}^{(i)}+\vect{n}_{t,f},\label{eq:obs}\\
\vect{x}_{t,f}^{(i)}&=\vect{d}_{t,f}^{(i)}+\vect{r}_{t,f}^{(i)},\label{eq:obs2}
\end{align}
where $t$ and $f$ are time and frequency indices, respectively, $\vect{x}_{t,f}=[x_{1,t,f},\ldots,x_{M,t,f}]^\top\in\mathbb{C}^{M\times 1}$ is a column vector containing all the microphone signals at a TF point. Here, $(\cdot)^{\top}$ denotes the non-conjugate transpose.
$\vect{x}_{t,f}^{(i)}=[x_{1,t,f}^{(i)},\ldots,x_{M,t,f}^{(i)}]^\top$ is a (noiseless) reverberant signal corresponding to the $i$th source, 
and $\vect{n}_{t,f}=[n_{1,t,f},\ldots,n_{M,t,f}]^\top$ is the additive diffuse noise.  
$\vect{x}_{t,f}^{(i)}$ for each source in Eq.~(\ref{eq:obs}) is further decomposed into two parts in Eq.~(2), one of which consists of the direct signal and early reflections, referred to as desired signal $\vect{d}_{t,f}^{(i)}$, and the other corresponds to late reverberation $\vect{r}_{t,f}^{(i)}$.
Hereafter, the frequency indices of the symbols are omitted for brevity, assuming that each frequency bin is processed independently in the same way. 

In this paper, the goal of {\DNDRSS} is to estimate $\vect{d}_t^{(i)}$ for each source $i$ from $\vect{x}_t$ in Eq.~(\ref{eq:obs}) by reducing $\vect{r}_t^{(i)}$ of source $i$, $\vect{x}_t^{(i')}$ of all the other sources $i'\neq i$, and diffuse noise $\vect{n}_t$. Since in noisy reverberant environments, early reflections enhance the intelligibility of speech for human perception \cite{early} and improve the ASR performance by computer \cite{Nishiura2007interspeech}, we include them in the desired signal.
Hereafter, we use $m=1$ as a reference microphone and describe a method for estimating desired signal $d_{1,t}^{(i)}$ at the microphone without loss of generality.

To achieve the above goal, we further model $\vect{d}_t^{(i)}$:
\begin{align}
\vect{d}_t^{(i)}&=\vect{v}^{(i)}s_t^{(i)}=\tilde{\vect{v}}^{(i)}d_{1,t}^{(i)},\label{eq:desired}
\end{align}
where $s_t^{(i)}$ is the $i$th clean speech at a TF point.
In Eq.~(\ref{eq:desired}), the desired signal of the $i$th source, $\vect{d}_t^{(i)}$, is modeled by $\vect{v}^{(i)}s_t^{(i)}$, i.e., a product in the STFT domain of the clean speech with transfer function $\vect{v}^{(i)}$, hereafter a steering vector, assuming that the duration of the impulse response corresponding to the direct signal and early reflections in the time domain is sufficiently short in comparison with the analysis window \cite{Avargel2007OnMT}.  
We further rewrite the desired signal as $\tilde{\vect{v}}^{(i)}d_{1,t}^{(i)}$, i.e., a product of the desired signal at reference microphone $d_{1,t}^{(i)}=v_1^{(i)}s_t^{(i)}$ with a Relative Transfer Function (RTF) \cite{RTF}, which is defined as the steering vector divided by its reference microphone element, 
\begin{align}
    \tilde{\vect{v}}^{(i)}={\vect{v}}^{(i)}/v_1^{(i)}.\label{eq:RTF}
\end{align}
In contrast, assuming that the duration of the late reverberation in the time domain exceeds the analysis window, late reverberation $\vect{r}_t^{(i)}$ is modeled by a convolution in the STFT domain \cite{Nak-ICASSP2008} of the clean speech with a time series of acoustic transfer functions that corresponds to the late reverberation:
\begin{align}
\vect{r}_t^{(i)}&=\sum_{\tau^={\Delta}}^{L_a-1} \vect{a}_{\tau}^{(i)}s_{t-\tau}^{(i)},\label{eq:late}
\end{align}
where $\vect{a}_{\tau}^{(i)}=[a_{1,\tau}^{(i)},\ldots,a_{M,\tau}^{(i)}]^\top$ for $\tau\in\{{\Delta},\ldots,L_a-1\}$ are the {convolutional} acoustic transfer functions, and ${\Delta}$ is {the mixing time, which represents} the relative frame delay of the late reverberation start time to the direct signal.


In this paper, we further assume that $\vect{d}_t^{(i)}$ is statistically independent\footnote{See a previous work \cite{wpe} for more precise discussion of the statistical independence between $\vect{d}_t^{(i)}$ and $s_{t'}$ for $t'\le t-\Delta$.} of {the following variables:
\begin{itemize}
    \item $s_{t'}^{(i)}$ for $t'\le t-\Delta$ (and thus $\vect{d}_t^{(i)}$ is statistically independent of $\vect{x}_{t'}^{(i)}$ for $t'\le t-\Delta$),
    \item $\vect{r}_{t''}^{(i)}$ for $t''\le t$,
    \item $\vect{x}_{t'}^{(i')}$ and $\vect{n}_{t'}$ for all $t$, $t'$ and $i'\neq i$.
\end{itemize}
These assumptions are used to derive the optimization algorithms described in the following.}

\subsection{Definition of a {\CBF} and its three different {\form}s}

We now define a {\CBF}, which will later br factorized into WPE filter(s) and beamformer(s):
\begin{align}
  \vect{y}_t&=\vect{W}_0^{\HT}\vect{x}_t+\sum_{\tau=\Delta}^{L-1}\vect{W}_{\tau}^{\HT}\vect{x}_{t-\tau},\label{eq:bf}
\end{align}
where $\vect{y}_t=[y_t^{(1)},\ldots,y_t^{(I)}]^{\top}\in\mathbb{C}^{I\times 1}$ is the output of the {\CBF} corresponding to the estimates of $I$ desired signals, $\vect{W}_{\tau}\in\mathbb{C}^{M\times I}$ for each $\tau\in\{0,\Delta,\Delta+1,\ldots,L-1\}$ is a matrix composed of the beamformer coefficients, $(\cdot)^{\HT}$ denotes a conjugate transpose, and $\Delta$ {is the prediction delay of {\CBF}. We set $\Delta$ equal to the mixing time introduced in Eq.~(\ref{eq:late}), so that the desired signals are included only in the first term of Eq.~(\ref{eq:bf}) and are statistically independent of the second term based on the assumptions introduced in the signal model. Then} this paper performs {\DNDRSS} by estimating the beamformer coefficients {that can estimate the desired signals included in the first term of} Eq.~(\ref{eq:bf}).  

{For notational simplicity, we also introduce a matrix representation of a {\CBF}:
\begin{align}
\vect{y}_t=\left[\begin{array}{c}\vect{W}_0\\\OL{\vect{W}}\end{array}\right]^{\HT}
\left[\begin{array}{c}\vect{x}_t\\\OL{\vect{x}}_t\end{array}\right],\label{eq:bfmtx}
\end{align}
where $\OL{\vect{W}}$ is a matrix containing $\vect{W}_\tau$ for $\Delta\le\tau\le L-1$ and $\OL{\vect{x}}_t$ is a column vector containing past multichannel observed signals $\vect{x}_{t-\tau}$ for $\Delta\le\tau\le L-1$:
\begin{align}
    \OL{\vect{W}}&=\left[\vect{W}_{\Delta}^{\top},\ldots,\vect{W}_{L-1}^{\top}\right]^{\top}\in\mathbb{C}^{M(L-\Delta)\times I},\\
    \OL{\vect{x}}_t&=\left[\vect{x}_{t-\Delta}^{\top},\ldots,\vect{x}_{t-L+1}^{\top}\right]^{\top}\in\mathbb{C}^{M(L-\Delta)\times 1}.
\end{align}
Hereafter, we refer to the {\CBF} defined by Eqs.~(\ref{eq:bf}) and (\ref{eq:bfmtx}) as a MIMO {\CBF}.} 

In the following, we further present three different {\form}s of {\CBF}, {including two ways of factorizing it. Figure~\ref{fig:imples} illustrates the MIMO {\CBF} and its three different implementations. }

\subsubsection{{\Mimo} factorization}
With the implementation shown in Fig.~\ref{fig:imples}~(b), we directly factorize\footnote{The existence of $\OL{\vect{G}}$, which satisfies $\OL{\vect{W}}=-\OL{\vect{G}}\vect{Q}$, is guaranteed for any $\OL{\vect{W}}$ when $M\ge I$ and $\text{rank}\{\vect{Q}\}=I$.}
the MIMO {\CBF} in Eq.~(\ref{eq:bfmtx}):
{\begin{align}
\left[\begin{array}{c}
\vect{W}_0\\\OL{\vect{W}}
\end{array}\right]=
\left[\begin{array}{c}\vect{I}_M\\-\OL{\vect{G}}\end{array}\right]\vect{Q},
\label{eq:factp}
\end{align}
where $\vect{Q}\in\mathbb{C}^{M\times I}$, $\OL{\vect{G}}\in\mathbb{C}^{M(L-\Delta)\times M}$, and $\vect{I}_M\in\mathbb{R}^{M\times M}$ is an identity matrix.}
Then Eq.~(\ref{eq:bf}) can be rewritten as a pair of a (convolutional) linear prediction filter followed by a (non-convolutional) beamformer matrix:
\begin{align}
    \vect{z}_t&=\vect{x}_t-\OL{\vect{G}}^{\HT}\OL{\vect{x}}_{t},\label{eq:bf_factorized1}\\
    \vect{y}_t&=\vect{Q}^{\HT}\vect{z}_t.\label{eq:bf_factorized}
\end{align}
Here $\vect{z}_t\in\mathbb{C}^{M\times 1}$ and $\OL{\vect{G}}$ are the output and the prediction matrix of the linear prediction, and $\vect{Q}$ is the coefficient matrix of the beamformer.
Eq.~(\ref{eq:bf_factorized1}), which is supposed to dereverberate all the sources at the same time, is thus referred to as a multiple-target linear prediction, and Eq.~(\ref{eq:bf_factorized}) is supposed to perform denoising and source separation at the same time. 
Because {individual sources are not distinguished in the WPE filter's output}, this implementation is called {\MIMO} factorization.

One example of {\MIMO} factorization is the cascade configuration composed of a WPE filter followed by a beamformer, which has been widely used for {\DNDRSS} in the far-field speech recognition area \cite{yoshioka2018interspeech,Kanda2019,NTTmeeting}, and the other example is one used in the joint optimization of a WPE filter and a beamformer, which has been investigated for DR+SS in the blind signal processing area  \cite{takuya2011taslp,ito2014iwaenc,Kagami2018icassp}.

\subsubsection{{Multi-Input Single-Output ({\TW}) {\CBF}}}
Next we define the set of {\tw} {\CBF}s shown in Fig.~\ref{fig:imples}~(c). They were obtained by decomposing {the beamformer coefficients in Eq.~(\ref{eq:bfmtx}):
\begin{align}
    \left[\begin{array}{c}
\vect{W}_0\\\OL{\vect{W}}
\end{array}\right]=
\left[\begin{array}{cccc}
\vect{w}_0^{(1)}&\vect{w}_0^{(2)}&\ldots&\vect{w}_0^{(I)}\\
\OL{\vect{w}}^{(1)}&\OL{\vect{w}}^{(2)}&\ldots&\OL{\vect{w}}^{(I)}
\end{array}\right],
\end{align}
where $\vect{w}_0^{(i)}\in\mathbb{C}^{M\times 1}$ and $\OL{\vect{w}}^{(i)}\in\mathbb{C}^{M(L-\Delta)\times 1}$ are column vectors, which respectively contain the $i$th columns of $\vect{W}_0$ and $\OL{\vect{W}}$; they are}
used to extract the $i$th desired signal.
Then, 
{Eq.~(\ref{eq:bfmtx})} can be rewritten for each source $i$:
\begin{align}
   y_t^{(i)}&=
   \left[\begin{array}{c}\vect{w}_0^{(i)}\\\OL{\vect{w}}^{(i)}\end{array}\right]^{\HT}
   \left[\begin{array}{c}\vect{x}_t\\\OL{\vect{x}}_t\end{array}\right].\label{eq:bf_sd}
\end{align}
%
For example, {\tw} CBFs were previously used  \cite{MLWPD,ikeshita2019waspaa}. ISCLP \cite{wpegsc} can also be viewed as the realization of a {\tw} CBF using a sidelobe cancellation framework \cite{onlinewpd}.

\subsubsection{{\Sw} factorization}
With the {\SW} factorization shown in Fig.~\ref{fig:imples}~(d), we further factorize each {\tw} {\CBF} defined in Eq.~(\ref{eq:bf_sd}) for source $i$:
{
\begin{align}
\left[\begin{array}{c}
\vect{w}_0^{(i)}\\\OL{\vect{w}}^{(i)}
\end{array}\right]=
\left[\begin{array}{c}\vect{I}_M\\-\OL{\vect{G}}^{(i)}\end{array}\right]\vect{q}^{(i)},\label{eq:factw}
\end{align}
where $\vect{q}^{(i)}\in\mathbb{C}^{M\times 1}$ and $\OL{\vect{G}}^{(i)}\in\mathbb{C}^{M(L-\Delta)\times M}$.
}
Then, Eq.~(\ref{eq:bf_sd}) can be rewritten as a pair of a linear prediction filter and a beamformer:
\begin{align}
    \vect{z}_t^{(i)}&=\vect{x}_t-\left(\OL{\vect{G}}^{(i)}\right)^{\HT}\OL{\vect{x}}_{t},\label{eq:sd_fact1}\\
    y_t^{(i)}&=\left(\vect{q}^{(i)}\right)^{\HT}\vect{z}_t^{(i)},\label{eq:sd_fact}
\end{align}
where $\vect{z}_t^{(i)}\in\mathbb{C}^{M\times 1}$ and $\OL{\vect{G}}^{(i)}$ are the output and the prediction matrix of the linear prediction, and $\vect{q}^{(i)}$ is the beamformer's coefficient vector.
Because Eq.~(\ref{eq:sd_fact1}) is performed only to estimate the $i$th source, it is called {\target} linear prediction.

\subsubsection{Relationship between two factorization approaches}
The difference between the two factorization approaches, namely Figs.~\ref{fig:imples}~(b) and (d), is based only on how the linear prediction is performed: Eq.~(\ref{eq:bf_factorized1}) or Eq.~(\ref{eq:sd_fact1}). More specifically, it is based on whether the prediction matrices, $\OL{\vect{G}}$ and $\OL{\vect{G}}^{(i)}$, 
are common to all the sources or different over different sources. Therefore, different optimization algorithms with different characteristics are derived, as will be shown in Section~\ref{sec:mlcb}. 
In contrast, the beamformer parts, $\vect{Q}$ and $\vect{q}^{(i)}$ in Eqs.~(\ref{eq:bf_factorized}) and (\ref{eq:sd_fact}) are identical in the two approaches, viewing $\vect{q}^{(i)}$ as the $i$th column of $\vect{Q}${, because they satisfy $\vect{W}_0=\vect{Q}$ in Eq.~(\ref{eq:factp}) and $\vect{w}_0^{(i)}=\vect{q}^{(i)}$ in Eq.~(\ref{eq:factw})}. 

In addition, it should be noted that all the above {\CBF} {\form}s are equivalent to each other in the sense that whatever values are set to the coefficients of one {\form}, certain coefficients of the other {\form}s can be determined such that they realize the same input-output relationship. Thus, the optimal solutions of all the {\form}s are identical as long as they are based on the same objective function.

\section{ML estimation of {\CBF}}\label{sec:mlcb}
In this section, we derive two different optimization algorithms using (b) {\MIMO} factorization and (d) {\SW} factorization. 
For the derivations, we assume that the RTFs $\tilde{\vect{v}}^{(i)}$ and the time-varying variances of the output signals yielded by the optimal {\CBF}, denoted by $\lambda_t^{(i)}$, are given. Then in Section~\ref{sec:flow}, we describe ways for jointly estimating $\lambda_t^{(i)}$ with {\CBF} coefficients based on the ML criterion and estimating $\tilde{\vect{v}}^{(i)}$ based on the WPE filter's output obtained at a step of the optimization.

\subsection{Probabilistic model}\label{sec:pmodel}
First, we formulate the objective function for {\DNDRSS} by reinterpreting the objective function proposed for DN+DR \cite{MLWPD}. 
For this formulation, we interpret {\DNDRSS} to be composed of a set of separate processing steps, each of which applies DN+DR to enhance source $i$ by
reducing the late reverberation of the source (DR) and the additive noise including the other sources and the diffuse noise (DN). 
With this interpretation, we introduce the following assumptions, similar to the previous work \cite{MLWPD}:
\begin{itemize}
    \item The output of the optimal {\CBF} for each $i$, namely ${y}_{t}^{(i)}$, follows a zero-mean complex Gaussian distribution with time-varying variance $\lambda_t^{(i)}=\mathbb{E}\left\{\left|{y}_{t}^{(i)}\right|^2\right\}$ \cite{wpe}.
    \item The beamformer satisfies a distortionless constraint for each source $i$ defined using RTF $\tilde{\vect{v}}^{(i)}$ in Eq.~(\ref{eq:RTF}):
    \begin{align}
        \left(\vect{w}_0^{(i)}\right)^{\HT}\tilde{\vect{v}}^{(i)}=1~\left(\mbox{or}~\left(\vect{q}^{(i)}\right)^{\HT}\tilde{\vect{v}}^{(i)}=1\right).\label{eq:dless}
    \end{align}
\end{itemize}
Then based on the previous discussion \cite{MLWPD}, we can approximately derive the objective function to be minimized for estimating the {\CBF} coefficients for source $i$, e.g., $\theta^{(i)}=\{\vect{w}_0^{(i)},\OL{\vect{w}}^{(i)}\}$, according to ML estimation:
\begin{align}
    {\cal L}_i(\theta^{(i)})&=\frac{1}{T}\sum_{t=1}^T\left(\frac{\left|y_{t}^{(i)}\right|^2}{\lambda_{t}^{(i)}}+\log\lambda_{t}^{(i)}\right)~\text{s.t.}~\left(\vect{w}_0^{(i)}\right)^{\HT}\tilde{\vect{v}}^{(i)}=1.\label{eq:sep_likelihood}
\end{align}
The objective function for estimating all the sources can then be obtained by summing Eq.~(\ref{eq:sep_likelihood}) over all the sources:
\begin{align}
    {\cal L}\left(\Theta\right)&=\sum_{i=1}^I{\cal L}_i(\theta^{(i)}),\label{eq:objective}
    ~\text{s.t.}~\left(\vect{w}_0^{(i)}\right)^{\HT}\tilde{\vect{v}}^{(i)}=1~\text{for all}~i,
\end{align}
where $\Theta=\left\{\theta^{(1)},\ldots,\theta^{(I)}\right\}$.
This objective function is used commonly for all the {\form}s of a {\CBF}.
{In this paper, we call a CBF optimized by the above objective function a weighted MPDR (wMPDR) CBF because it minimizes the average power of output $y_t^{(i)}$ weighted by time-varying variance, $\lambda_t^{(i)}$, of the signal.}

Here, let us briefly explain how {\DNDRSS} is performed by Eqs.~(\ref{eq:sep_likelihood}) and (\ref{eq:objective}). Substituting Eqs.~(\ref{eq:obs}) and (\ref{eq:obs2}) into Eq.~(\ref{eq:bf_sd}) and using the model of the desired signal in Eq.~(\ref{eq:desired}) and the distortionless constraint in Eq.~(\ref{eq:dless}), we obtain
\begin{align}
   y_t^{(i)}&={d}_{1,t}^{(i)}+\hat{r}_t^{(i)}+\sum_{i'\neq i}\hat{x}_{t}^{(i')}+\hat{n}_t,\label{eq:apply}%
\end{align}
where $\hat{r}_t^{(i)}$, $\hat{x}_t^{(i')}$ for $i'\neq i$, and $\hat{n}_t$ are respectively the late reverberation of the $i$th source, all the other sources, and the additive diffuse noise remaining in the {\CBF} output, written in {\tw} {\CBF} form:
{\begin{align}
    \hat{r}_t^{(i)}&=
    \left[\begin{array}{c}\vect{w}_0^{(i)}\\\OL{\vect{w}}^{(i)}\end{array}\right]^{\HT}
    \left[\begin{array}{c}\vect{r}_t^{(i)}\\\OL{\vect{x}}_t^{(i)}\end{array}\right],\label{eq:rhat}\\
    \hat{x}_t^{(i')}&=
    \left[\begin{array}{c}\vect{w}_0^{(i)}\\\OL{\vect{w}}^{(i)}\end{array}\right]^{\HT}
    \left[\begin{array}{c}\vect{x}_t^{(i')}\\\OL{\vect{x}}_t^{(i')}\end{array}\right],\label{eq:xhat}\\
    \hat{n}_t&=
    \left[\begin{array}{c}\vect{w}_0^{(i)}\\\OL{\vect{w}}^{(i)}\end{array}\right]^{\HT}
    \left[\begin{array}{c}\vect{n}_t\\\OL{\vect{n}}_t\end{array}\right],\label{eq:nhat}
\end{align}
where $\OL{\vect{n}}_t=[\vect{n}_{t-\Delta}^{\top},\ldots,\vect{n}_{t-L+1}^{\top}]^{\top}$.}
According to the statistical independence assumptions introduced in Section~\ref{sec:model}, $d_{1,t}^{(i)}$ is statistically independent of $\hat{r}_t^{(i)}$, $\hat{x}_t^{(i')}$, and $\hat{n}_t$. Then substituting Eq.~(\ref{eq:apply}) into Eq.~(\ref{eq:sep_likelihood}) and omitting the constant terms, we obtain the following (in the expectation sense):
\begin{align}
    \mathbb{E}\left\{{\cal L}_i(\theta^{(i)})\right\}&=\frac{1}{T}\sum_{t=1}^T\frac{\mathbb{E}\left\{\left|\hat{r}_t^{(i)}+\sum_{i'\neq i}\hat{x}_{t}^{(i')}+\hat{n}_t\right|^2\right\}}{\lambda_{t}^{(i)}}.
\end{align}
The above equation indicates that minimization of the objective function indeed minimizes the sum of $\hat{r}_t^{(i)}$, $\hat{x}_t^{(i')}$ for $i'\neq i$, and $\hat{n}_t$ in Eq.~(\ref{eq:apply}).

Before deriving the optimization algorithms, we define a matrix that is frequently used in the derivation, referred to as a variance-normalized spatio-temporal covariance matrix.  Letting $\LOL{\vect{x}}_t$ be a column vector composed of the current and past observed signals at all the microphones, defined as
\begin{align}
    \LOL{\vect{x}}_t=\left[\vect{x}_t^{\top}, \OL{\vect{x}}_t^{\top}\right]^{\top}\in\mathbb{C}^{M(L-\Delta+1)\times 1}, \label{eq:longx}
\end{align}
the matrix is defined:
\begin{align}
    \LOL{\vect{R}}_{\vect{x}}^{(i)}=\frac{1}{T}\sum_{t=1}^{T}\frac{\LOL{\vect{x}}_t\LOL{\vect{x}}_t^{\HT}}{\lambda_t^{(i)}}\in\mathbb{C}^{M(L-\Delta+1)\times M(L-\Delta+1)}.\label{eq:tscov}
\end{align}
Its factorized form is also defined:
\begin{align}
    \LOL{\vect{R}}_{\vect{x}}^{(i)}=\left[\begin{array}{cc}
    \vect{R}_{\vect{x}}^{(i)}&\left(\vect{P}_{\vect{x}}^{(i)}\right)^{\HT}\\
    \vect{P}_{\vect{x}}^{(i)}&\OL{\vect{R}}_{\vect{x}}^{(i)}
    \end{array}\right],\label{eq:tscov_fact}
\end{align}
where
\begin{align}
    \vect{R}_{\vect{x}}^{(i)}&=\frac{1}{T}\sum_{t=1}^T\frac{{\vect{x}}_t{\vect{x}}_t^{\HT}}{\lambda_t^{(i)}}\in\mathbb{C}^{M\times M},\label{eq:tscov_fact1}\\
    \vect{P}_{\vect{x}}^{(i)}&=\frac{1}{T}\sum_{t=1}^T\frac{\OL{\vect{x}}_t{\vect{x}}_t^{\HT}}{\lambda_t^{(i)}}\in\mathbb{C}^{M(L-\Delta)\times M},\label{eq:tscov_fact2}\\ \OL{\vect{R}}_{\vect{x}}^{(i)}&=\frac{1}{T}\sum_{t=1}^T\frac{\OL{\vect{x}}_t\OL{\vect{x}}_t^{\HT}}{\lambda_t^{(i)}}\in\mathbb{C}^{M(L-\Delta)\times M(L-\Delta)}.\label{eq:tscov_fact3}
\end{align}

\subsection{Optimization based on {\MIMO} factorization}\label{opt_mb}
This subsection discusses methods for optimizing a {\CBF} with the {\MIMO} factorization. In the following, after describing a method for directly applying the conventional joint optimization technique used for DR+SS to {\DNDRSS}, we summarize the problems in it, and present the solutions to the problems.

\subsubsection{Direct application of a conventional technique}\label{sec:simple}
With the {\MIMO} factorization in Eqs.~(\ref{eq:bf_factorized1}) and (\ref{eq:bf_factorized}), simultaneously estimating both $\vect{Q}$ and $\OL{\vect{G}}$ in closed form is difficult even when both $\lambda_t^{(i)}$ and $\tilde{\vect{v}}^{(i)}$ are given. Instead, we use an iterative and alternate estimation scheme, following a blind signal processing technique \cite{takuya2011taslp,ito2014iwaenc,Kagami2018icassp}, where at each estimation step, either $\vect{Q}$ or $\OL{\vect{G}}$ is updated and the other is fixed. 

For updating $\OL{\vect{G}}$, we fix $\vect Q$ at its previously estimated value. For the algorithm derivation, the representation of linear prediction in Eq.~(\ref{eq:bf_factorized1}) is slightly modified:
\begin{align}
    \vect{z}_t&=\vect{x}_{t}-\OL{\vect{X}}_{t}\OL{\vect{g}},\label{eq:gbar}
\end{align}
where $\OL{\vect{X}}_{t}$ and $\OL{\vect{g}}$ are equivalent to $\OL{\vect{x}}_{t}$ and $\OL{\vect{G}}$ with a modified matrix structure defined:
\begin{align}
    \OL{\vect{X}}_{t}&=\vect{I}_M\otimes\OL{\vect{x}}_{t}^{\top}
    \in\mathbb{C}^{M\times M^2(L-\Delta)},\label{eq:Y1}\\
    \OL{\vect{g}}&=\left[\OL{\vect{g}}_{1}^{\top},\ldots,\OL{\vect{g}}_{M}^{\top}\right]^{\HT}\in\mathbb{C}^{M^2(L-\Delta)\times 1},\label{eq:g2}
\end{align}
where $\otimes$ is a Kronecker product and $\OL{\vect{g}}_{m}$ is the $m$th column of $\OL{\vect{G}}$.
Then, considering that the {\CBF} in Eqs.~(\ref{eq:bf_factorized1}) and (\ref{eq:bf_factorized}) can be written as ${y}_t^{(i)}=\left(\vect{q}^{(i)}\right)^{\HT}\left(\vect{x}_{t}-\OL{\vect{X}}_{t}\OL{\vect{g}}\right)$ and omitting the normalization terms, the objective function in Eq.~(\ref{eq:objective}) becomes
\begin{align}
    {\cal L}_{\OL{\vect{g}}}(\OL{\vect{g}})&=\frac{1}{T}\sum_{t=1}^T\left\|\vect{x}_{t}-\OL{\vect{X}}_{t}\OL{\vect{g}}\right\|_{\vect{\Phi}_{\vect{q},t}}^2,\label{eq:GQ}
\end{align}
where $\left\|\vect{x}\right\|_{\vect{R}}^2=\vect{x}^{\HT}\vect{R}\vect{x}$, and $\vect{\Phi}_{\vect{q},t}$ is a semi-definite Hermitian matrix:
\begin{align}
    \vect{\Phi}_{\vect{q},t}&=\sum_{i=1}^I\frac{\vect{q}^{(i)}\left(\vect{q}^{(i)}\right)^{\HT}}{\lambda_{t}^{(i)}} \in\mathbb{C}^{M\times M}.\label{eq:Rq}
\end{align}
Because Eq.~(\ref{eq:GQ}) is a quadratic form with a lower bound, $\OL{\vect{g}}$, which minimizes it, can be obtained:
\begin{align}
    \OL{\vect{g}}&=\vect{\Psi}^{+}{\boldsymbol\psi},\label{eq:MS_wpe1}\\
    {{\vect{\Psi}}}&=\frac{1}{T}\sum_t\OL{\vect{X}}_{t}^{\HT}\vect{\Phi}_{\vect{q},t}\OL{\vect{X}}_{t}\in\mathbb{C}^{M^2(L-\Delta)\times M^2(L-\Delta)},\label{eq:MS_wpe2}\\
    {\boldsymbol\psi}&=\frac{1}{T}\sum_t\OL{\vect{X}}_{t}^{\HT}\vect{\Phi}_{\vect{q},t}\vect{x}_{t}\in\mathbb{C}^{M^2(L-\Delta)\times 1},\label{eq:MS_wpe3}
\end{align}
where $(\cdot)^+$ is the Moore-Penrose pseudo-inverse.  
Since the rank of $\vect\Psi$ is equal to or smaller than $MI(L-\Delta)$, as shown in Section~\ref{sec:advanced}, $\vect\Psi$ is rank deficient for over-determined cases, namely when $M>I$, and thus the use of the pseudo-inverse is indispensable. 
Eqs.~(\ref{eq:MS_wpe1}) to (\ref{eq:MS_wpe3}) are equivalent to those used in the dereverberation step for DR+SS \cite{takuya2011taslp,ito2014iwaenc,Kagami2018icassp} except that in our paper denoising is additionally included in the objective and over-determined cases are also considered.  We call this a multiple-target WPE filter.

For the update of $\vect{Q}$, fixing $\OL{\vect{g}}$ at its previously estimated value, the objective in Eq.~(\ref{eq:objective}) can be rewritten:
\begin{align}
    {\cal L}_{\vect{Q}}(\vect{Q})=\sum_{i=1}^I\left\|\vect{q}^{(i)}\right\|_{\vect{R}_{\vect{z}}^{(i)}}^2~\text{s.t.}~\left(\vect{q}^{(i)}\right)^{\HT}\tilde{\vect{v}}^{(i)}=1,\label{eq:qlikelihood}
\end{align}
where $\vect{R}_{\vect{z}}^{(i)}$ is a variance-normalized spatial covariance matrix of the output of the multiple-target WPE filter, calculated as
\begin{align}
    \vect{R}_{\vect{z}}^{(i)}&=\frac{1}{T}\sum_{t=1}^T\frac{{\vect{z}}_{t}{\vect{z}}_{t}^{\HT}}{\lambda_{t}^{(i)}}.\label{eq:Rnacute}
\end{align}
Then $\vect{q}^{(i)}$, which minimizes Eq.~(\ref{eq:qlikelihood}) under the distortionless constraint $\left(\vect{q}^{(i)}\right)^{\HT}\tilde{\vect{v}}^{(i)}=1$, can be obtained:
\begin{align}
    \vect{q}^{(i)}&=\frac{\left(\vect{R}_{\vect{z}}^{(i)}\right)^{-1}\tilde{\vect{v}}^{(i)}}{\left(\tilde{\vect{v}}^{(i)}\right)^{\HT}\left(\vect{R}_{\vect{z}}^{(i)}\right)^{-1}\tilde{\vect{v}}^{(i)}}.\label{eq:qblock}
\end{align}
Because the above beamformer minimizes the average power of $\vect{z}_t$ weighted by the time-varying variance, we call it a weighted MPDR (wMPDR) beamformer\footnote{A wMPDR beamformer was also called a Maximum-Likelihood Distortionless Response (MLDR) beamformer \cite{Cho2019MLDR}.}. {As shown in Section~\ref{sec:swcbf}, a wMPDR beamformer is a special case of a wMPDR {\CBF}, which is reduced to a wMPDR beamformer when setting the length of the {\CBF} $L=1$, i.e., by just converting it into a non-convolutional beamformer.}

The above algorithm, however, has two serious problems. First, the size of the covariance matrix in Eq.~(\ref{eq:MS_wpe2}) is too large, requiring huge computing cost for calculating it and its inverse. Second, as shown in our experiments, the iterative and alternate estimation of $\vect{Q}$ and $\OL{\vect{G}}$ tends to converge to a sub-optimal point. 
This is probably because 
the update of $\OL{\vect{G}}$ is performed based only on the output of the fixed beamformer in the iterative and alternate estimation, as in Eq.~(\ref{eq:sep_likelihood}); the signal dimension of the beamformer output, i.e., $I$, is reduced from that of the original signal space, i.e., $M$, with the over-determined case, i.e., $I<M$. As a consequence, signal
%
components that are relevant for the update of $\OL{\vect{G}}$ may be reduced in the beamformer output, especially when the estimation of $\vect{Q}$ is less accurate at the early stage of the optimization. This can seriously degrade the update of $\OL{\vect{G}}$. 
%

\subsubsection{Proposed extension}\label{sec:advanced}
Next we present two techniques to mitigate the above problems within the {\MIMO} factorization approach. The first reduces the computing cost.  As shown in Appendix~\ref{sec:Psifast}, Eqs.~(\ref{eq:MS_wpe2}) and (\ref{eq:MS_wpe3}) can be rewritten, using Eq.~(\ref{eq:tscov_fact}):
\begin{align}
    {{\vect{\Psi}}}&=\sum_{i=1}^I\left(\vect{q}^{(i)}\left(\vect{q}^{(i)}\right)^{\HT}\otimes\left(\OL{\vect{R}}_{\vect{x}}^{(i)}\right)^{\top}\right),\label{eq:Psifast}\\
    {\boldsymbol\psi}&=\sum_{i=1}^I\left(\vect{q}^{(i)}\otimes\left(\vect{P}_{\vect{x}}^{(i)}\vect{q}^{(i)}\right)^{*}\right),\label{eq:psifast}
\end{align}
where $()^{*}$ denotes the complex conjugate. In the above equations, the majority of the calculation is coming from $\OL{\vect{R}}_{\vect{x}}^{(i)}$. Because the size of the matrix is much smaller than that of $\vect\Psi$, we can greatly reduce the computing cost with this modification\footnote{{In general, the computational complexity  of a matrix multiplication exceeds $O(n^2)$. Because the size of $\Psi$ is $M$-times larger than $\OL{\vect{R}}_{\vect{x}}$, the computational complexity for calculating $\vect{\Psi}$ is probably at least $M^2$ times larger than that for calculating $\OL{\vect{R}}_{\vect{x}}$.}} in comparison with the direct calculation of Eqs.~(\ref{eq:MS_wpe2}) and (\ref{eq:MS_wpe3}). Although we still need to calculate the inverse of huge matrix $\vect\Psi$ even with this modification, the cost is relatively small in comparison with the direct calculation of $\vect\Psi$. Note that Eq.~(\ref{eq:Psifast}) also shows the rank of $\vect{\Psi}$ to be equal to or smaller than $MI(L-\Delta)$.

The second technique introduces a heuristic to improve the update of the WPE filter. {To use a whole $M$-dimensional signal space to be considered for the update,} we modify the {\CBF} to output not only $I$ desired signals, but also $M-I$ auxiliary signals that are included in orthogonal complement $\vect{Q}^{\perp}$ of $\vect{Q}$ and model the auxiliary signals as zero-mean time-varying complex Gaussians. 
With this modification, the optimization is performed by calculating the summation in Eqs.~(\ref{eq:Psifast}) and (\ref{eq:psifast}) over both $1\le i \le I$ and $I<i\le M$, letting $\vect{q}^{(I+1)},\ldots,\vect{q}^{(M)}$ be the orthonormal bases for the orthogonal complement $\vect{Q}^{\perp}$.
Because distinguishing variances $\lambda_t^{(i)}$ of the auxiliary signals is inconsequential, we use the same value for them, calculated as 
\begin{align}
  {\lambda_t^{\perp}=\frac{1}{M-I}\sum_{i=I+1}^M\left|\left(\vect{q}^{(i)}\right)^{\HT}\vect{z}_t\right|^2,}
\end{align}
and calculate ${\vect{P}}_{\vect{x}}^{\perp}$ and $\OL{\vect{R}}_{\vect{x}}^{\perp}$ based on Eqs.~(\ref{eq:tscov_fact2}) and (\ref{eq:tscov_fact3}) accordingly. In summary, we can implement this modification by adding the following terms to $\vect\Psi$ and $\boldsymbol\psi$ in Eqs.~(\ref{eq:Psifast}) and (\ref{eq:psifast}):
\begin{align}
    \vect{\Psi}^{\perp}&=\left(\sum_{i=I+1}^M \vect{q}^{(i)}\left(\vect{q}^{(i)}\right)^{\HT}\right)\otimes\left(\OL{\vect{R}}_{\vect{x}}^{\perp}\right)^{\top},\\
    {\boldsymbol\psi}^{\perp}&=\sum_{i=i+1}^M\left(\vect{q}^{(i)}\otimes\left(\vect{P}_{\vect{x}}^{\perp}\vect{q}^{(i)}\right)^{*}\right).
\end{align}

\subsection{Direct optimization of {\tw} {\CBF}s}\label{sec:swcbf}
Before deriving the optimization with {\SW} factorization, we show that we can directly optimize the {\tw} {\CBF}s in Eq.~(\ref{eq:bf_sd}), and summarize their characteristics.
With this setting, the {\CBF}s and the objective function are both defined separately for each source in Eqs.~(\ref{eq:bf_sd})~and~ (\ref{eq:sep_likelihood}), and thus, the optimization can be performed separately for each source. 
%
The resultant algorithm is, therefore, identical to that previously proposed for DN+DR \cite{wpdarxiv}, {where this type of {\CBF} is also called a Weighted Power minimization Distortionless response (WPD) {\CBF}}. 

For presenting the solution, we introduce the following vector representation of Eq.~(\ref{eq:bf_sd}):
\begin{align}
y_t^{(i)}=\left(\LOL{\vect{w}}^{(i)}\right)^{\HT}\LOL{\vect{x}}_t,\label{eq:unifsol}
\end{align}
where $\LOL{\vect{w}}^{(i)}$ is defined:
\begin{align}
    \LOL{\vect{w}}^{(i)}&=\left[\begin{array}{c}
    \vect{w}_0^{(i)}\\\OL{\vect{w}}^{(i)}\end{array}\right],
\end{align}
Then, when $\lambda_t^{(i)}$ and $\tilde{\vect{v}}^{(i)}$ are given, Eq.~(\ref{eq:sep_likelihood}) becomes a simple constraint quadratic form:
\begin{align}
{\cal L}_{\vect{w}}(\LOL{\vect{w}}^{(i)})=\left\|\LOL{\vect{w}}^{(i)}\right\|_{\LOL{\vect{R}}_{\vect{x}}^{(i)}}^2~\text{s.t.}~\left(\LOL{\vect{w}}^{(i)}\right)^{\HT}\LOL{\vect{v}}^{(i)}=1,
\end{align}
where $\LOL{\vect{R}}_{\vect{x}}^{(i)}$ is the covariance matrix defined in Eq.~(\ref{eq:tscov_fact}), and $\LOL{\vect{v}}^{(i)}=\left[\left(\tilde{\vect{v}}^{(i)}\right)^{\top},0,\ldots,0\right]^{\top}\in\mathbb{C}^{M(L-\Delta+1)\times 1}$ corresponds to the RTF $\tilde{\vect{v}}^{(i)}$ with zero padding.
Finally, we obtain the solution:
\begin{align}
    \LOL{\vect{w}}^{(i)}=\frac{\left(\LOL{\vect{R}}_{\vect{x}}^{(i)}\right)^{-1}\LOL{\vect{v}}^{(i)}}{\left(\LOL{\vect{v}}^{(i)}\right)^{\HT}\left(\LOL{\vect{R}}_{\vect{x}}^{(i)}\right)^{-1}\LOL{\vect{v}}^{(i)}}.\label{eq:unified_sol}
\end{align}
{The above equation, which gives the simplest form of the solution to a wMPDR {\CBF}, clearly shows that a wMPDR {\CBF} is a general case of a wMPDR beamformer. By setting $L=1$ in the above solution, namely, by letting it be a non-convolutional beamformer, it reduces to the solution of a wMPDR beamformer in Eq.~(\ref{eq:qblock}).}

An advantage of the solution using the {\tw} {\CBF}s is that it can be obtained by a closed form equation, provided the RTFs and the time-varying variances of the desired signals are given and that we can ignore the interaction between DN and DR. With this approach, however, the RTFs must be directly estimated from a reverberant observation, similar to ISCLP \cite{wpegsc}. A solution to this problem is to use dereverberation preprocessing based on a WPE filter for the RTF estimation. Although it was shown that the output of a WPE filter can be obtained in a computationally efficient way within the framework of this approach \cite{MLWPD}, the {\SW} factorization approach described in the following can more naturally solve this problem. So, this paper adopts it as the solution.

\subsection{Optimization based on {\SW} factorization}\label{opt_sw}
With {\SW} factorization, similar to the case with the direct optimization of the {\tw} {\CBF}s, the optimization can be performed separately for each source, and the resultant algorithm is identical to that proposed for DN+DR  \cite{factorizedwpd}.

Considering that a {\CBF} can be written based on Eqs.~(\ref{eq:sd_fact1}) and  (\ref{eq:sd_fact}) as $y_t^{(i)}=\left(\vect{q}^{(i)}\right)^{\HT}\left(\vect{x}_t-\left(\OL{\vect{G}}^{(i)}\right)^{\HT}\OL{\vect{x}}_t\right)$  and using the factorized form of $\LOL{\vect{R}}_{\vect{x}}^{(i)}$ in Eq.~(\ref{eq:tscov_fact}), the objective function in Eq.~(\ref{eq:sep_likelihood}) can be rewritten:
\begin{align}
    {\cal L}_i\left(\OL{\vect{G}}^{(i)},\vect{q}^{(i)}\right)=&\left\|\left(\OL{\vect{G}}^{(i)}-\left(\OL{\vect{R}}_{\vect{x}}^{(i)}\right)^{-1}\vect{P}_{\vect{x}}^{(i)}\right)\vect{q}^{(i)}\right\|_{\OL{\vect{R}}_{\vect{x}}^{(i)}}^2\nonumber\\
    &~\hspace{-10mm}+\left\|\vect{q}^{(i)}\right\|_{\left(\vect{R}_{\vect{x}}^{(i)}-\left(\vect{P}_{\vect{x}}^{(i)}\right)^{\HT}\left(\OL{\vect{R}}_{\vect{x}}^{(i)}\right)^{-1}\vect{P}_{\vect{x}}^{(i)}\right)}^2.\label{eq:objective_Gq2} 
\end{align}
In the above objective function, $\OL{\vect{G}}^{(i)}$ is contained only in the first term, and the term can be minimized without depending on the value of $\vect{q}^{(i)}$, when $\OL{\vect{G}}^{(i)}$ takes the following value:
\begin{align}
    \OL{\vect{G}}^{(i)}=\left(\OL{\vect{R}}_{\vect{x}}^{(i)}\right)^{-1}\vect{P}_{\vect{x}}^{(i)}.\label{eq:wpesol}
\end{align}
So, this is a solution\footnote{This is not a unique solution. The first term is minimized even when an arbitrary matrix, whose null space includes $\vect{q}^{(i)}$, is added to Eq.~(\ref{eq:wpesol}).}
of $\OL{\vect{G}}^{(i)}$ that globally minimizes the objective function given time-varing variance $\lambda_t^{(i)}$. Interestingly, this solution is identical to that of conventional WPE dereverberation.  This means that the WPE filter, which is optimized solely for dereverberation, can perform the optimal dereverberation for the joint optimization without depending on the subsequent beamforming, provided the time-varying variance of the desired source is given for the optimization. In addition, unlike the {\MIMO} factorization approach, this approach does not need to compensate for the {dimensionality reduction of the beamformer output for the update of $\OL{\vect{G}}^{(i)}$ because it considers a whole signal space without adding any modification.}
We refer to this filter $\OL{\vect{G}}^{(i)}$ as a {\target} WPE filter.
\begin{algorithm}[t!]
\SetAlgoLined
\DontPrintSemicolon
\AlgoDontDisplayBlockMarkers\SetAlgoNoEnd\SetAlgoNoLine
 \KwData{Observed signal $\vect{x}_t$ for all $t$\;
 \hspace{0.96cm}TF masks $\gamma_t^{(i)}$ for all $t$ and $1\le i\le I$}
 \KwResult{Estimated sources $y_{t}^{(i)}$ for all $t$ and $1\le i \le I$}
 Initialize $\lambda_t^{(i)}$ as $||\vect{x}_{t}||_{\vect{I}_M}^2/M$ for all $t$ and $1\le i\le I$\;
 Initialize $\vect{q}^{(i)}$ as the $i$th column of $\vect{I}_M$ for $1\le i\le I$\;
 Initialize $\vect{z}_t$ as $\vect{x}_t$ for all $t$\;
 \Repeat{convergence}{
  $\OL{\vect{R}}_{\vect{x}}^{(i)}\leftarrow\frac{1}{T}\sum_{t=1}^T\frac{\OL{\vect{x}}_t\OL{\vect{x}}_t^{\HT}}{\lambda_t^{(i)}}$ for $1\le i\le I$\;
  ${\vect{P}}_{\vect{x}}^{(i)}\leftarrow\frac{1}{T}\sum_{t=1}^T\frac{\OL{\vect{x}}_t{\vect{x}}_t^{\HT}}{\lambda_t^{(i)}}$ for $1\le i\le I$\;
  ${\vect{\Psi}}\leftarrow\sum_{i=1}^I\left( \vect{q}^{(i)}\left(\vect{q}^{(i)}\right)^{\HT}\otimes\left(\OL{\vect{R}}_{\vect{x}}^{(i)}\right)^{\top}\right)$\;
  ${{\boldsymbol\psi}}\leftarrow\sum_{i=1}^I\left(\vect{q}^{(i)}\otimes\left(\vect{P}_{\vect{x}}^{(i)}\vect{q}^{(i)}\right)^{*}\right)$\;
\textbf{Begin} Add orthogonal complement beamformer\;
  \Indp Set $\vect{q}^{(I+1)},\ldots,\vect{q}^{(M)}$ as the orthonormal bases for orthogonal complement $\vect{Q}^{\perp}$ of $\vect{Q}$\;
  {$\lambda_t^{\perp}\leftarrow\frac{1}{M-I}\sum_{i=I+1}^M\left|\left(\vect{q}^{(i)}\right)^{\HT}\vect{z}_t\right|^2$} \;
  $\OL{\vect{R}}_{\vect{x}}^{\perp}\leftarrow\frac{1}{T}\sum_{t=1}^T\frac{\OL{\vect{x}}_t\OL{\vect{x}}_t^{\HT}}{\lambda_t^{\perp}}$\;
  ${\vect{P}}_{\vect{x}}^{\perp}\leftarrow\frac{1}{T}\sum_{t=1}^T\frac{\OL{\vect{x}}_t{\vect{x}}_t^{\HT}}{\lambda_t^{\perp}}$\;
  ${\vect{\Psi}}\leftarrow{\vect{\Psi}}+\left(\sum_{i=I+1}^M \vect{q}^{(i)}\left(\vect{q}^{(i)}\right)^{\HT}\right)\otimes\left(\OL{\vect{R}}_{\vect{x}}^{\perp}\right)^{\top}$\;
  ${\boldsymbol\psi}\leftarrow{\boldsymbol\psi}+\sum_{i=I+1}^M\left(\vect{q}^{(i)}\otimes\left(\vect{P}_{\vect{x}}^{\perp}\vect{q}^{(i)}\right)^{*}\right)$\;
  \Indm
\textbf{End}\;%
  $\OL{\vect{g}}\leftarrow\vect{\Psi}^{+}{\boldsymbol\psi}$\;
  $\vect{z}_t\leftarrow\vect{x}_{t}-\OL{\vect{X}}_{t}\OL{\vect{g}}$\; 
  Estimate $\tilde{\vect{v}}^{(i)}$ based on $\vect{z}_t$ and $\gamma_t^{(i)}$ for $1\le i\le I$\;
  ${\vect{R}}_{\vect{z}}^{(i)}\leftarrow\frac{1}{T}\sum_{t=1}^T\frac{\vect{z}_t\left(\vect{z}_t\right)^{\HT}}{\lambda_t^{(i)}}$ for $1\le i\le I$\;
  $\vect{q}^{(i)}\leftarrow\frac{\left({\vect{R}}_{\vect{z}}^{(i)}\right)^{+}\tilde{\vect{v}}^{(i)}}{\left(\tilde{\vect{v}}^{(i)}\right)^{\HT}\left({\vect{R}}_{\vect{z}}^{(i)}\right)^{+}\tilde{\vect{v}}^{(i)}}$ for $1\le i\le I$\;
  $y_t^{(i)}\leftarrow\left(\vect{q}^{(i)}\right)^{\HT}\vect{z}_t$ for $1\le i\le I$\;
  $\lambda_t^{(i)}\leftarrow\left|{y}_t^{(i)}\right|^2$ for $1\le i\le I$\;
 }
 \caption{{\Mimo} factorization-based optimization for estimation of all sources}\label{alg:mb}
\end{algorithm}

\begin{algorithm}[t!]
\SetAlgoLined
\DontPrintSemicolon
\AlgoDontDisplayBlockMarkers\SetAlgoNoEnd\SetAlgoNoLine%
 \KwData{Observed signal $\vect{x}_t$ for all $t$\;
 \hspace{0.96cm}TF masks $\gamma_t^{(i)}$ for all $t$}
 \KwResult{Estimated $i$th source $y_{t}^{(i)}$ for all $t$}
 Initialize $\lambda_t^{(i)}$ as $||\vect{x}_{t}||_{\vect{I}_M}^2/M$ for all $t$\;
 \Repeat{convergence}{
  $\OL{\vect{R}}_{\vect{x}}^{(i)}\leftarrow\frac{1}{T}\sum_{t=1}^T\frac{\OL{\vect{x}}_t\OL{\vect{x}}_t^{\HT}}{\lambda_t^{(i)}}$\;
  ${\vect{P}}_{\vect{x}}^{(i)}\leftarrow\frac{1}{T}\sum_{t=1}^T\frac{\OL{\vect{x}}_t{\vect{x}}_t^{\HT}}{\lambda_t^{(i)}}$\;
  $\OL{\vect{G}}^{(i)}\leftarrow\left(\OL{\vect{R}}_{\vect{x}}^{(i)}\right)^{+}\vect{P}_{\vect{x}}^{(i)}$\;
  $\vect{z}_t^{(i)}\leftarrow\vect{x}_t-\left(\OL{\vect{G}}^{(i)}\right)^{\HT}\OL{\vect{x}}_t$\;
  Estimate $\tilde{\vect{v}}^{(i)}$ based on $\vect{z}_t^{(i)}$ and $\gamma_t^{(i)}$\;
  $\acute{\vect{R}}_{\vect{z}}^{(i)}\leftarrow\frac{1}{T}\sum_{t=1}^T\frac{\vect{z}_t^{(i)}\left(\vect{z}_t^{(i)}\right)^{\HT}}{\lambda_t^{(i)}}$\;
  $\vect{q}^{(i)}\leftarrow\frac{\left(\acute{\vect{R}}_{\vect{z}}^{(i)}\right)^{+}\tilde{\vect{v}}^{(i)}}{\left(\tilde{\vect{v}}^{(i)}\right)^{\HT}\left(\acute{\vect{R}}_{\vect{z}}^{(i)}\right)^{+}\tilde{\vect{v}}^{(i)}}$\;
  $y_t^{(i)}\leftarrow\left(\vect{q}^{(i)}\right)^{\HT}\vect{z}_t^{(i)}$\;
  $\lambda_t^{(i)}\leftarrow\left|{y}_t^{(i)}\right|^2$\;
 }
 \caption{{\Sw} factorization-based optimization for estimation of $i$th source}\label{alg:sw}
\end{algorithm}

Once $\OL{\vect{G}}^{(i)}$ is obtained as the above solution, the objective function in Eq.~(\ref{eq:sep_likelihood}) can be rewritten as
\begin{align}
    {\cal L}_i\left(\vect{q}^{(i)}\right)=\left\|\vect{q}^{(i)}\right\|_{\acute{\vect{R}}_{\vect{z}}^{(i)}}^2~\text{s.t.}~\left(\vect{q}^{(i)}\right)^{\HT}\tilde{\vect{v}}^{(i)}=1,
    \label{eq:objective_q}
\end{align}
where $\acute{\vect{R}}_{\vect{z}}^{(i)}$ is a variance-normalized covariance  matrix of the output of the {\target} WPE filter, calculated as
\begin{align}
    \acute{\vect{R}}_{\vect{z}}^{(i)}&=\frac{1}{T}\sum_{t=1}^T\frac{{\vect{z}}_{t}^{(i)}\left({\vect{z}}_{t}^{(i)}\right)^{\HT}}{\lambda_{t}^{(i)}}\in\mathbb{C}^{M\times M}.\label{eq:Racute}
\end{align}
Then the solution can be obtained, under a distortionless constraint, as a wMPDR beamformer:
\begin{align}
    \vect{q}^{(i)}&=\frac{\left(\acute{\vect{R}}_{\vect{z}}^{(i)}\right)^{-1}\tilde{\vect{v}}^{(i)}}{\left(\tilde{\vect{v}}^{(i)}\right)^{\HT}\left(\acute{\vect{R}}_{\vect{z}}^{(i)}\right)^{-1}\tilde{\vect{v}}^{(i)}}.\label{eq:qfact}
\end{align}
Eqs.~(\ref{eq:objective_q}) to (\ref{eq:qfact}) closely resemble Eqs.~(\ref{eq:qlikelihood}) to (\ref{eq:qblock}). The difference is whether the dereverberation is performed by a multiple-target WPE filter or {\target} WPE filters.

With {\SW} factorization, the solution can be obtained in closed form when $\lambda_t^{(i)}$ and $\tilde{\vect{v}}^{(i)}$ are given, similar to the case with the direct optimization of the {\tw} {\CBF}s. In addition, the output of the WPE filter is obtained as $\vect{z}_t^{(i)}$ in Eq.~(\ref{eq:sd_fact1}), and can be efficiently used for the estimation of the RTFs. Furthermore, since the temporal-spatial covariance matrix in Eq.~(\ref{eq:tscov_fact3}) is much smaller than that in Eq.~(\ref{eq:MS_wpe2}) of the {\MIMO} factorization, the computational cost can be reduced. (See Section~\ref{sec:discussion} for more scrutiny of the computing cost.) 

\subsection{Processing flow with estimation of $\lambda_t^{(i)}$ and  $\tilde{\vect{v}}^{(i)}$}\label{sec:flow}
This subsection describes examples of processing flows in Algorithms~\ref{alg:mb} and \ref{alg:sw}, for optimizing a {\CBF} based on {\MIMO} factorization and {\SW} factorization, including estimation of the time-varying variances, $\lambda_t^{(i)}$, and the RTFs, $\tilde{\vect{v}}^{(i)}$. 
Hereafter, we refer to the algorithms as \mbox{A-1} and \mbox{A-2} for brevity.
Although \mbox{A-1} simultaneously estimates all sources, $y_t^{(i)}$ for all $i$, from observed signal $\vect{x}_t$, \mbox{A-2} estimates only one of the sources, $y_t^{(i)}$ for a certain $i$, and (if necessary) is repeatedly applied to the observed signal to estimate all the sources one after another. 
TF masks are provided as auxiliary inputs for both algorithms.
TF mask $\gamma_t^{(i)}$, which is associated with a source and a TF point, takes a value between 0 and 1 and indicates whether the source's desired signal dominates the TF point ($\gamma_t^{(i)}=1$) or not ($\gamma_t^{(i)}=0$). The TF masks over all the TF points are used to estimate the RTF(s) of the desired signal(s) in line 19 of \mbox{A-1} and line 7 of \mbox{A-2}.  (See Section~\ref{sec:TFRTF} for the estimation detail of the TF masks and the RTFs.)

Both algorithms estimate time-varying variances $\lambda_t^{(i)}$ based on the same objective as that for the {\CBF}, defined in Eq.~(\ref{eq:sep_likelihood}). Because no closed form solution to the estimation of the {\CBF} and the time-varying variances is known, an iterative and alternate optimization scheme is introduced to both algorithms. In each iteration, the time-varying variances, $\lambda_t^{(i)}$, are updated in line 23 of \mbox{A-1} and line 11 of \mbox{A-2} as the power of the previously estimated values of desired signal $y_t^{(i)}$, and then the {\CBF} and desired signal $y_t^{(i)}$ are updated while fixing the time-varying variances.  The iteration is repeated until convergence is obtained.

The optimization methods described in Sections~\ref{opt_mb} and \ref{opt_sw} are used in their respective algorithms to update the {\CBF} and the desired signal(s). 
The WPE filter is first estimated in lines 5 to 17 of \mbox{A-1} and lines 3 to 5 of \mbox{A-2}, and applied in line 18 of \mbox{A-1} and line 6 of \mbox{A-2}. After the RTF(s) is updated using the dereverberated signals, the wMPDR beamformer is estimated in lines 20 and 21 of \mbox{A-1} and lines 8 and 9 of \mbox{A-2}, and applied in line 22 of \mbox{A-1} and line 10 of \mbox{A-2}.

\begin{figure}[t]
\centering
\includegraphics[width=0.9\columnwidth]{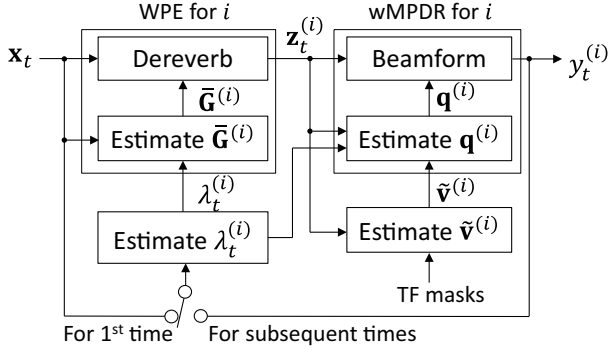}
\caption{Processing flow of {\SW} factorization-based {\CBF} for estimating a source $i$.}\label{fig:flow}
\end{figure}

Figure~\ref{fig:flow} also illustrates the processing flow of a {\CBF} with {\SW} factorization for estimating a source $i$.

\begin{table*}
\centering
\caption{{{\CBF}s compared in experiments: (1) and (2) are conventional cascade configuration approaches, (5) is a conventional joint optimization approach, (6) and (7) are proposed joint optimization approaches, and (3) and (4) are test conditions used just for comparison.  (5), (6), and (7) are categorized as ``Jointly optimal'' because they are composed of WPE and wMPDR and optimized based on integrated variance estimation (see Fig.~\ref{fig:sepvsjoin} for the difference between separate and integrated variance estimation).}}\label{fig:treediag}
\begin{tabular}{lcccccc}\toprule
 Name of method & Jointly & WPE & BF & Variance & Category\\
 & optimal & & & estimation\\\midrule
 (1) WPE+MPDR (separate) & & Multiple-target & MPDR & Separate & Cascade (conventional)\\
 (2) WPE+MVDR (separate) & & Multiple-target & MVDR & Separate & Cascade (conventional)\\
 (3) WPE+wMPDR (separate) & & Multiple-target & wMPDR & Separate & Test condition\\
 (4) WPE+MPDR (integrated) & & Single-target & MPDR & Integrated & Test condition\\\hline
(5) Source-packed factorization (conventional) & \checkmark & Multiple-target & wMPDR & Integrated & Jointly optimal (conventional)\\
 (6) Source-packed factorization (extended) & \checkmark & Multiple-target & wMPDR & Integrated & Jointly optimal (proposed)\\
 (7) Source-wise factorization & \checkmark & Single-target & wMPDR & Integrated & Jointly optimal (proposed)\\\bottomrule
\end{tabular}
\end{table*}

\subsubsection{Methods for estimating TF masks and RTFs}\label{sec:TFRTF}
In our experiments, for estimating TF masks, $\gamma_t^{(i)}$, for all $i$ and $t$ {at each frequency}, we used a Convolutional Neural Network {that works in the TF domain} and is trained using utterance-level Permutation Invariant Training criterion (CNN-uPIT) \cite{CNN-uPIT}.  According to our preliminary experiments \cite{nak2020icassp}, we set the network structure as a CNN with a large receptive field similar to one used by a fully-Convolutional Time-domain Audio Separation Network (Conv-TasNet) \cite{ConvTasnet2019TASLP}. 
%
The network was trained so that it receives the WPE filter's output, which is obtained at the first iteration in the iterative optimization of the {\CBF}, and estimates the TF masks of the desired signals. The network's input was set as a concatenation of the real and imaginary parts of the STFT coefficients, and the loss function was set as the (scale-dependent) signal-to-distortion ratio (SDR) of an enhanced signal obtained by multiplying the estimated masks to an observed signal. For the training and validation data, we synthesized mixtures using two utterances randomly extracted from the WSJ-CAM0 corpus \cite{wsjcam0} and two room impulse responses and background noise extracted from the REVERB Challenge training set \cite{REVERB}.

For the estimation of the RTFs, $\tilde{\vect{v}}^{(i)}$, we adopted a method based on eigenvalue decomposition with noise covariance whitening \cite{ito17icassp,MGolan09taslp}.  With this technique, steering vector $\vect{v}^{(i)}$ is first estimated:
\begin{align}
    \vect{v}^{(i)}={\cal R}_{\setminus i}\mbox{MaxEig}\left({\cal R}_{\setminus i}^{-1}{\cal R}_{i}\right),
\end{align}
where $\mbox{MaxEig}(\cdot)$ is a function that calculates the eigenvector corresponding to the maximum eigenvalue and ${\cal R}_{i}$ and ${\cal R}_{{\setminus}i}$ are spatial covariance matrices of the $i$-th desired signal and the other signals estimated as:
\begin{align}
    {\cal R}_{i}&=\frac{\sum_t\gamma_{t}^{(i)}\vect{z}_t^{(i)}\left(\vect{z}_t^{(i)}\right)^{\HT}}{\sum_t\gamma_{t}^{(i)}},\\
    {\cal R}_{{\setminus}i}&=\frac{\sum_t\left(1-\gamma_{t}^{(i)}\right)\vect{z}_t^{(i)}\left(\vect{z}_t^{(i)}\right)^{\HT}}{\sum_t\left(1-\gamma_{t}^{(i)}\right)}.
\end{align}
Then, the RTF is obtained by Eq.~(\ref{eq:RTF}).

\section{Discussion}\label{sec:discussion}
In summary, our proposed techniques can optimize a {\CBF} for jointly performing {\DNDRSS} with greatly reduced computing cost in comparison with the direct application of the conventional joint optimization technique proposed for DR+SS to {\DNDRSS}.  
With the conventional technique, a huge covariance matrix $\vect\Psi$ must be calculated to take into account the dependency of $\OL{\vect{G}}$ on $\vect{Q}$ that is inherently introduced into {\MIMO} factorization. This makes the computing cost of the conventional technique extremely high. In contrast, since the proposed extension of the {\MIMO} factorization approach substantively reduces the size of the matrix to be calculated from $M^2(L-\Delta)$ for $\vect\Psi$ to $M(L-\Delta)$ for $\OL{\vect{R}}_{\vect{x}}$, the computing cost can be effectively reduced. 

On the other hand, with {\SW} factorization, $\OL{\vect{G}}^{(i)}$ can be optimized independently of $\vect{q}^{(i)}$, which also allows us to reduce the size of the matrix to be calculated to the same as that of the proposed extension of the {\MIMO} factorization approach. In addition, we can skip the calculation of an additional matrix, $\OL{\vect{R}}_{\vect{x}}^{\perp}$, {and the inverse of the huge matrix, $\vect{\Psi}^{-1}$}, both of which are required for the proposed extension of the {\MIMO} factorization approach. This further increases the computational efficiency of the {\SW} factorization approach. A drawback of {\SW} factorization is that it has to handle $I$-times more dereverberated signals than {\MIMO} factorization.

The {\SW} factorization approach has additional benefits w.r.t. computational efficiency when it is used in specific scenarios listed below:
\begin{itemize}
\item The {\SW} factorization approach can estimate the {\CBF} by a closed-form equation when time-varying source variances are given, or estimated, e.g., using neural networks \cite{YongXu,DNN-WPE}. In such a case, we can skip iterative optimization. In contrast, the {\MIMO} factorization approach needs to maintain iterations to alternately estimate $\vect{Q}$ and $\OL{\vect{g}}$ due to their mutual dependency.
\item The {\SW} factorization approach is advantageous when it is combined with neural network-based single target speaker extraction that has recently been actively studied \cite{speakerbeam}. With this combination, we can skip the estimation of sources other than the target source, allowing us to further reduce the computing cost.
\end{itemize}

\section{Experiments}\label{sec:exp}
This section experimentally confirms the effectiveness of our proposed joint optimization approaches. 
{Table~\ref{fig:treediag}} summarizes the optimization methods that we experimentally compared (see Sections~\ref{sec:exp1} and \ref{sec:exp2} for details of the methods) in the following three aspects.
\begin{enumerate}
\item {Effectiveness of joint optimization} \\
We compared a {\CBF} with and without joint optimization in terms of estimation accuracy. The {\SW} factorization approach ({Table~\ref{fig:treediag}~(7)}) is compared with the conventional cascade configuration ({Table~\ref{fig:treediag}~(1) and (2)}), and two additional test conditions ({Table~\ref{fig:treediag}~(3) and (4)}).
\item {Comparison among joint optimization approaches} \\
We compared three joint optimization approaches, i.e., the {\MIMO} factorization approach with its conventional setting ({Table~\ref{fig:treediag}~(5)}) and its proposed extension ({Table~\ref{fig:treediag}~(6)}), and the {\SW} factorization approach ({Table~\ref{fig:treediag}~(7)}), respectively described in Sections~\ref{sec:simple}, \ref{sec:advanced}, and \ref{opt_sw}, in terms of computational efficiency and estimation accuracy. 
\item {Evaluation using oracle masks}\\
{We used oracle masks instead of estimated masks for evaluating a {\CBF} to test the performance of a {\CBF} using different types of masks and also to obtain its top-line performance. }
\end{enumerate}

\subsection{Dataset and evaluation metrics}
\label{conditions}
\begin{figure}[t]
\centering
\includegraphics[width=0.9\columnwidth]{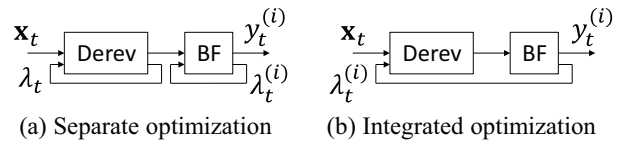}
\caption{Separate and integrated variance optimization schemes: {While separate variance optimization updates $\lambda_t$ for Derev as the variance of Derev output, integrated variance optimization updates it as the variance of the beamformer output. Consequently, $\lambda_t$ for Derev is common to all the sources with separate variance optimization.}}\label{fig:sepvsjoin}
\end{figure}

For the evaluation, we prepared a set of noisy reverberant speech mixtures (REVERB-2MIX) using the REVERB Challenge dataset (REVERB) \cite{REVERB}. Each utterance in REVERB contains a single reverberant speech with moderate stationary diffuse noise.  For generating a set of test data, we mixed two utterances extracted from REVERB, one from its development set (Dev set) and the other from its evaluation set (Eval set), so that each pair of mixed utterances was recorded in the same room, by the same microphone array, and under the same condition (near or far, RealData or SimData). We categorized the test data based on the original categories of the data in REVERB (e.g., SimData or RealData).  We created the same number of mixtures in the test data as in the REVERB Eval set, such that each utterance in the REVERB Eval set is contained in one of the mixtures in the test data. Furthermore, the length of each mixture in the test data was set at the same as that of the corresponding utterance in the REVERB Eval set, {and the utterance from the Dev set was trimmed or zero-padded at its end to be the same length as that of Eval set.}

{For the experiments in Section~\ref{sec:exp3}, we also prepared a set of noisy reverberant speech mixtures, each of which is composed of three speaker utterances (REVERB-3MIX). We created REVERB-3MIX by adding one utterance extracted from REVERB Dev set to each mixture in REVERB-2MIX. Only RealData (i.e., real recordings of reverberant data) was created for REVERB-3MIX. }

In the experiments, we respectively estimated two or three speech signals from each mixture for REVERB-2MIX and REVERB-3MIX and evaluated only one of them corresponding to the REVERB Eval set using the baseline evaluation tools provided for it. We selected the signal to be evaluated from all the estimated speech signals based on the correlation between the separated signals and the original signal in the REVERB Eval set. As objective measures for speech enhancement \cite{metrics}, we used the Cepstrum Distance (CD), the Frequency-Weighted Segmental SNR (FWSSNR), the Perceptual Evaluation of Speech Quality (PESQ), and {the Short-Time Objective Intelligibility measure (STOI) \cite{stoi}}. To evaluate the ASR performance, we used a baseline ASR system for REVERB that was recently developed using Kaldi \cite{kaldi}. This system is composed of a Time-Delay Neural Network (TDNN) acoustic model trained using lattice-free maximum mutual information (LF-MMI) and online i-vector extraction, and a trigram language model. They were trained on the REVERB training set. 

\subsection{{\CBF} configurations}



Table~I summarizes two configurations of the CBF examined in experiments including the number of microphones $M$, the filter length $L$, and the number of optimization iterations.
The sampling frequency was 16 kHz. A Hann window was used for a short-time analysis where the frame length and shift were set at 32 and 8 ms. The prediction delay was set at $\Delta = 4$ for the WPE filter. 
\begin{table}[t]
\renewcommand{\arraystretch}{1.1}
\centering
\caption{Beamformer configurations used in experiments}
\begin{tabular}{cccccc}\hline
& $M$ & \multicolumn{3}{c}{$L$ at each freq.\ range (kHz)}  &\#Iterations\\\cline{3-5}
& & 0.0-0.8 & 0.8-1.5 & 1.5-8.0 & \\\hline
Config-1 & 8 & 20 & 16 & 8 & 10\\
Config-2 & 4 & 20 & 16 & 8 & 10\\\hline
\end{tabular}
\end{table}

In the iterative optimization, the time-varying variances of the sources were initialized as those of the observed signal for the WPE filter and as 1 for the wMPDR beamformer for all the methods.

\subsection{Experiment-1: effectiveness of joint optimization}\label{sec:exp1}
\label{results}

\begin{table}[!t]
\renewcommand{\arraystretch}{1.1}
\caption{WER (\%) for RealData and CD (dB), FWSSNR (dB), PESQ, and STOI for SimData in REVERB-2MIX obtained using different beamformers after five estimation iterations with Config-1. Scores for REVERB-2MIX and REVERB (i.e., single speaker) without enhancement (No Enh), are also shown.}\label{tbl:aq0}
\centering
\begin{tabular}{lp{6mm}p{4.5mm}p{6mm}p{6mm}p{5.5mm}}
\toprule
\mcc{Enhancement method}  & {WER} & \mcc{CD} & 
\multicolumn{1}{p{7mm}}{~\hspace{-2mm}FWSSNR} & {PESQ} & {STOI}\\
\midrule
No Enh (REVERB-2MIX) & 62.49 & 5.44 & ~~~1.12 & ~1.12 & {~0.55}\\
No Enh (REVERB) & 18.61 & 3.97 & ~~~3.62 & ~1.48 & {~0.75}\\
\midrule
MPDR (w/o iteration) & 30.79 & 4.40 & ~~~3.07 & ~1.45 & {~0.73}\\
{MVDR (w/o iteration)} & {30.89} & {4.43} & ~~~{3.00} & ~{1.44} &~{0.73}\\
wMPDR & 28.75 & 3.96 & ~~~4.46 & ~1.60 & {~0.75}\\
\midrule
(1) WPE+MPDR (separate) & 23.04 & 4.30 & ~~~3.77 & ~1.58 & {~0.77}\\
{(2) WPE+MVDR (separate)} & {23.34} & {4.34} & ~~~{3.66} & ~{1.57} & ~{0.76}\\
(3) WPE+wMPDR (separate) & 21.53 & 3.74 & ~~~5.42 & ~1.77 & ~{{\bf 0.82}}\\
(4) WPE+MPDR (integrated) & 23.22 & 4.28 & ~~~3.66 & ~1.56 & {~0.76}\\
(7) Source-wise factorization & {\bf 20.03} & {\bf 3.67} & ~~~{\bf 5.57} & ~{\bf 1.80} & {~{0.81}}\\
\bottomrule
\end{tabular}
\end{table}
\begin{figure}[t]
\centering
\includegraphics[width=0.96\columnwidth]{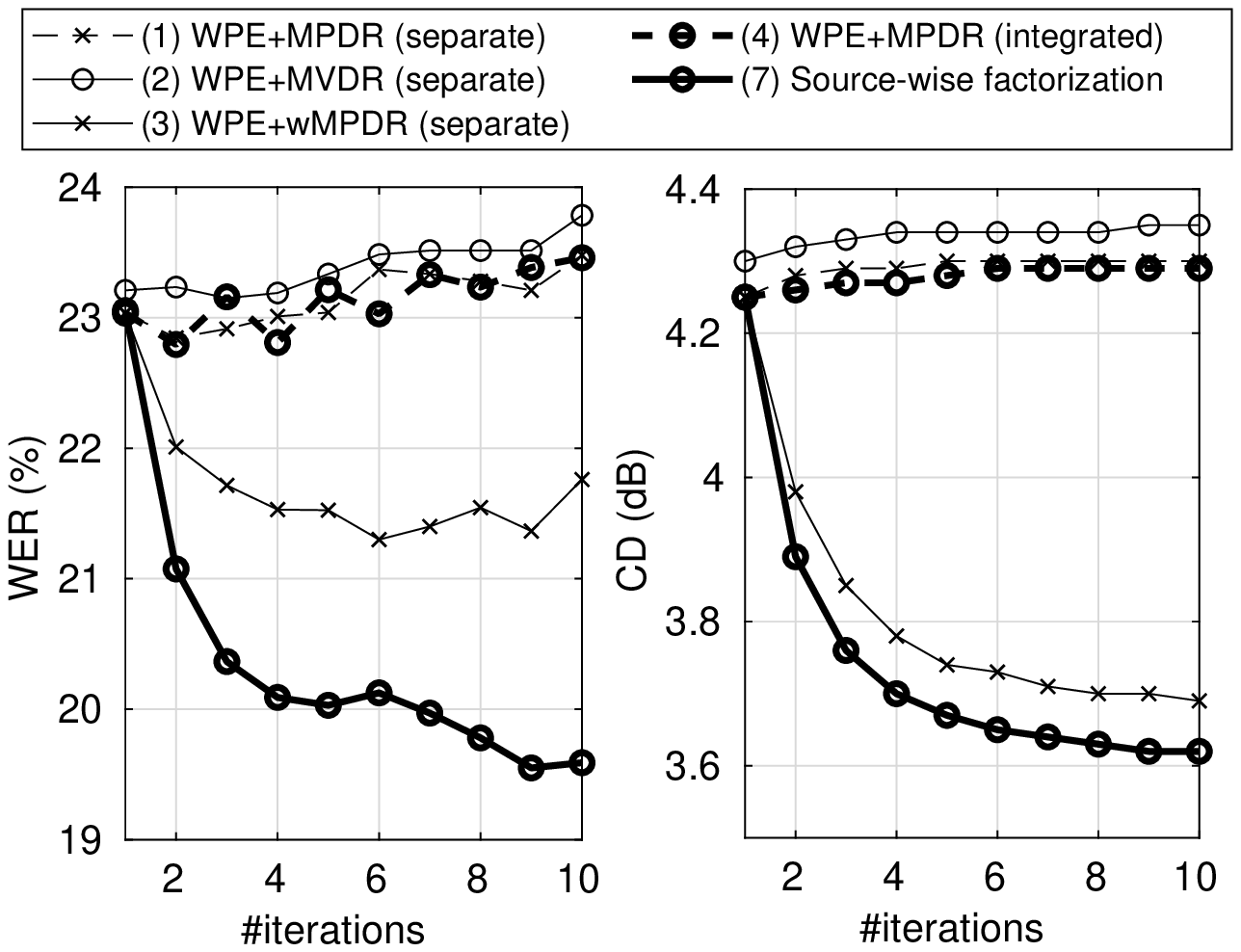}\\
~\hspace{-4mm}\includegraphics[width=0.96\columnwidth]{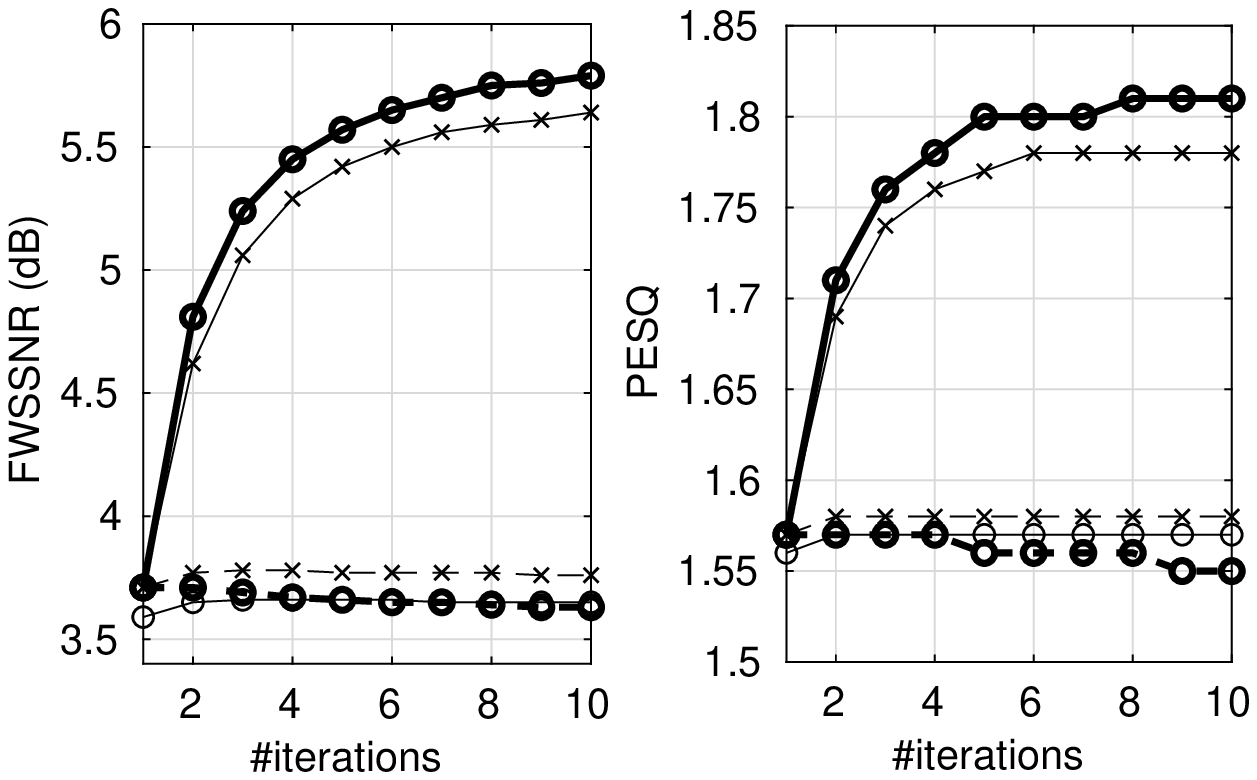}
\caption{Comparison among joint optimization and cascade configuration approaches when using WPE+MPDR and WPE+wMPDR with integrated and separate optimization schemed using Config-1 for REVERB-2MIX.} \label{exp:sepvsjoint}
\end{figure}
\begin{figure*}[t]
\centering
\begin{tabular}{cc}
 \includegraphics[width=8cm]{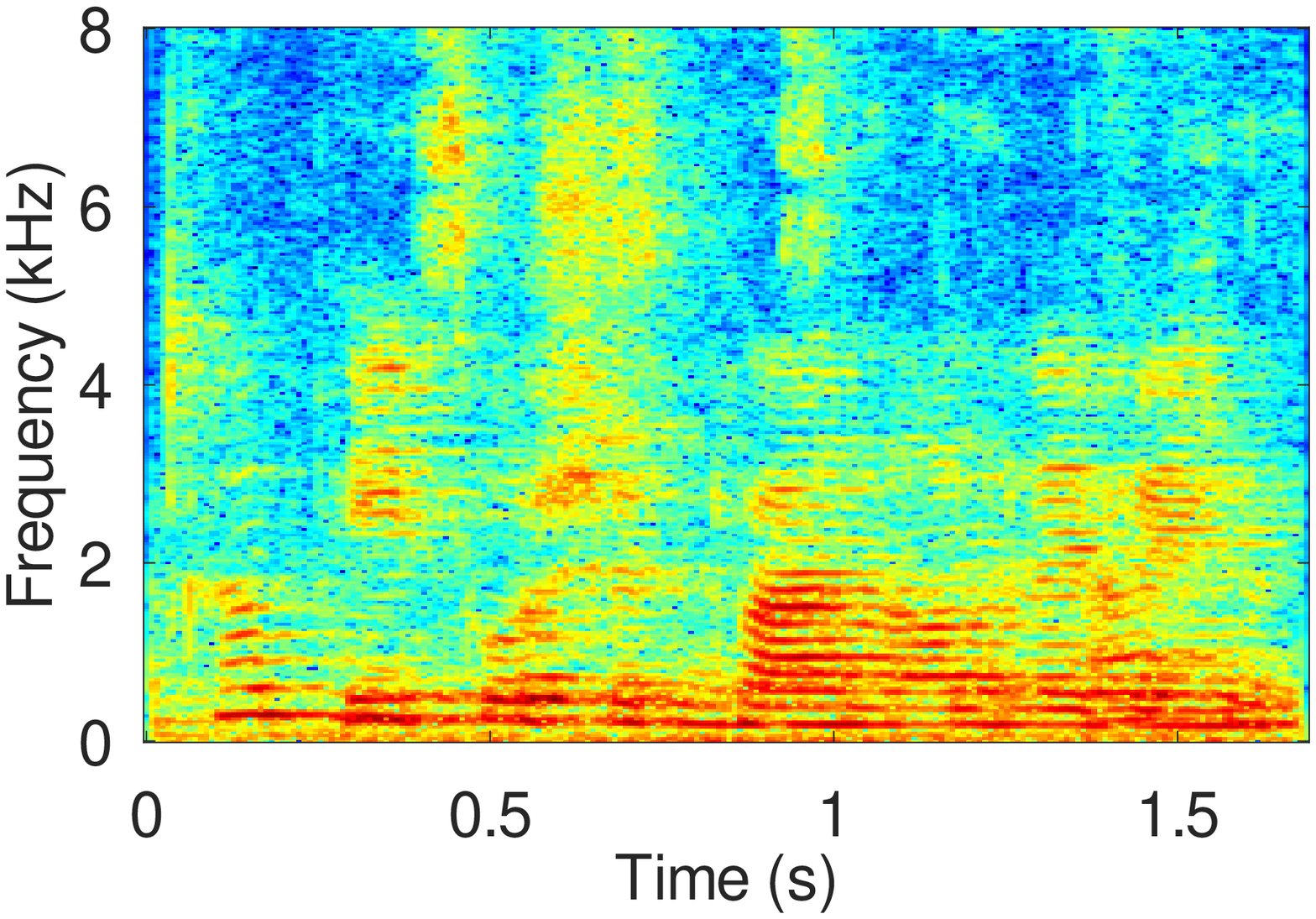}& \includegraphics[width=8cm]{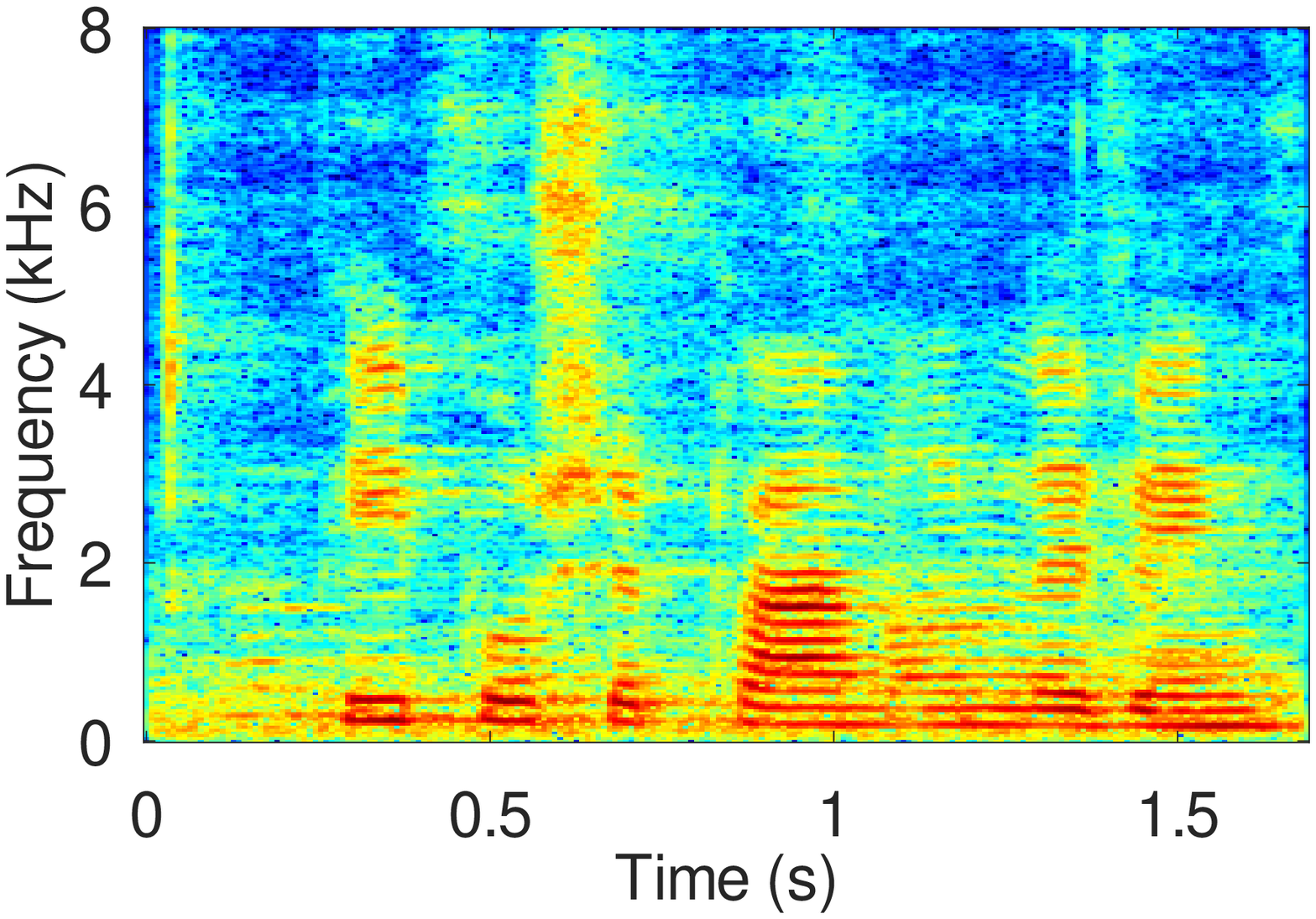}\\
 (a) Observed signal& (b) MVDR\\
\includegraphics[width=8cm]{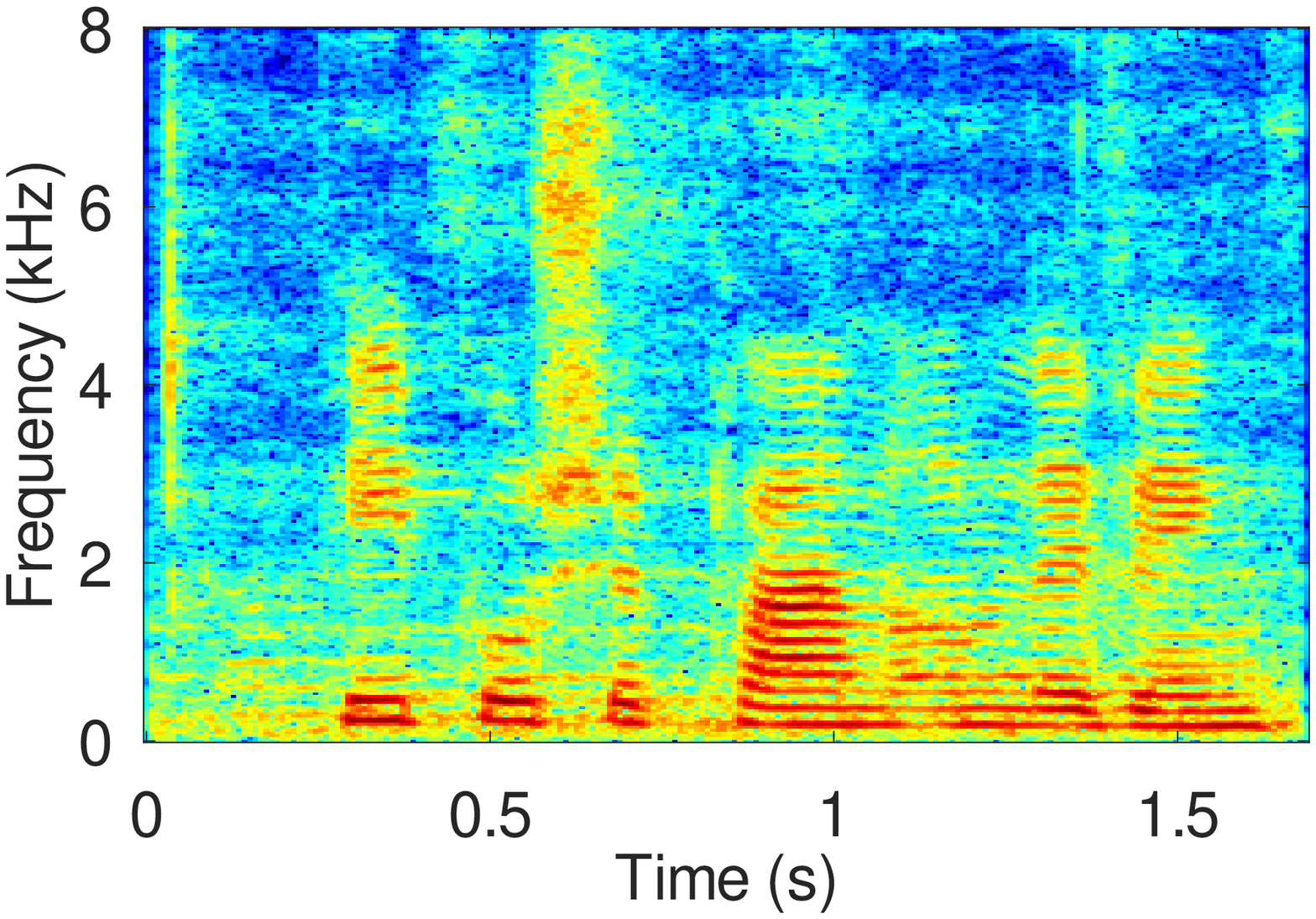}& \includegraphics[width=8cm]{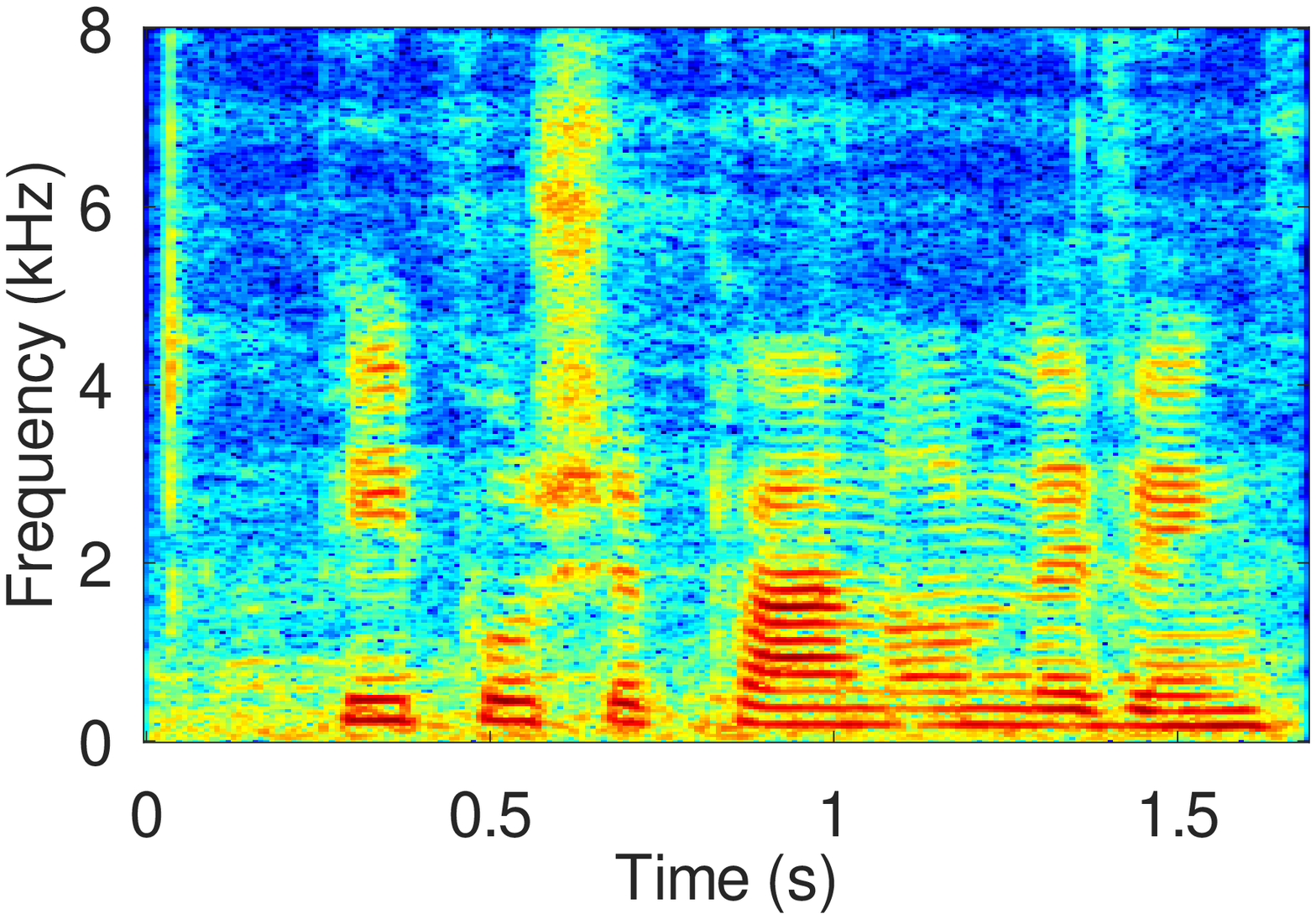}\\
(c) WPE+MVDR & (d) CBF with source-wise factorization
 \end{tabular}
 \caption{{Spectrogram of (a) a noisy reverberant mixture in RealData of REVERB-2MIX and spectrograms of enhanced signals obtained by (b) MVDR, (c) WPE+MVDR and (d) CBF with source-wise factorization. Mixture is composed of two female speakers under far conditions.}}\label{fig:spec}
\end{figure*}
In this experiment, we evaluated the effectiveness of the joint optimization focusing on its two characteristics. First, we compared {three} different filter combinations: a WPE filter followed by a wMPDR beamformer (WPE+wMPDR), a WPE filter followed by an MPDR beamformer (WPE+MPDR){, and a WPE filter followed by an MVDR beamformer (WPE+MVDR)}. The first combination is required for jointly optimal processing, and the others have been used for the conventional cascade configuration. Second, we compared two different variance optimization schemes shown in Fig.~\ref{fig:sepvsjoin}: ``separate'' and ``integrated.'' With the separate variance optimization, the iterative estimation of the time-varying variance was performed separately for the WPE filter and for the beamformer.  This is the scheme used by the conventional cascade configuration. In contrast, with the integrated variance optimization, the iterative estimation was performed jointly for the WPE filter and the beamformer. A significant difference between the two schemes is whether the WPE filter uses the same variances for all the sources or different variances dependent on the sources estimated by the beamformer. 

Table~\ref{tbl:aq0} compares WERs, CDs, FWSSNRs, PESQs, {and STOIs} obtained after five estimation iterations using three beamformers (MPDR, {MVDR,} and wMPDR), two conventional cascade configuration approaches ((1) WPE+MPDR and {(2) WPE+MVDR}), two test conditions ((3) and (4)), and a proposed joint optimization approach ((7) {\SW} factorization). All methods used configuration Config-1 in Table~I. 
Table~\ref{tbl:aq0} shows that 1)~WPE+MPDR, {WPE+MVDR}, and WPE+wMPDR greatly outperformed MPDR, {MVDR}, and wMPDR, respectively, with all the conditions, 2) the joint optimization approach, i.e., (7) source-wise factorization, substantially outperformed all the other methods in terms of all the measures {except for a case in terms of STOI where WPE+wMPDR (separate) gave a slightly better score than (7) source-wise factorization}. 
Furthermore, Fig.~\ref{exp:sepvsjoint} shows the convergence curves of the two cascade configuration approaches, two test conditions, and the joint optimization approach. The source-wise factorization performance (7) was the best of all and improved as the number of iterations increased. The second best was (3) WPE+wMPDR (separate). The other methods did not improve the scores after the first iteration with both the integrated and separate variance optimization schemes.

{Figure~\ref{fig:spec} shows a spectrogram of a noisy reverberant mixture in RealData of REVERB-2MIX, and spectrograms of enhanced signals obtained using MVDR, WPE+MVDR, and CBF with source-wise factorization. The figure shows that all the enhancement methods were effective and the CBF with source-wise factorization was the best of all for achieving denoising, dereverberation, and source separation. }

The above results clearly show that the two characteristics of the joint optimization approach, i.e., 1) the optimal combination of a WPE filter and a wMPDR beamformer, and 2) the integrated variance optimization, are both critical for achieving optimal performance.

\subsection{Experiment-2: Comparison among joint optimization approaches}\label{sec:exp2}

In this experiment, we compared three joint optimization approaches, denoted as (5) {\Mimo} factorization (conventional), (6) {\Mimo} factorization (extended), and (7) {\Sw} factorization. 
(5) {\Mimo} factorization (conventional) corresponds to the conventional joint optimization technique described in Section~\ref{sec:simple}, and (6) {\Mimo} factorization (extended) and (7) {\Sw} factorization correspond to our proposed methods respectively described in Sections~\ref{sec:advanced} and \ref{opt_sw}. 
\begin{figure}[t]
\centering
\includegraphics[width=0.95\columnwidth]{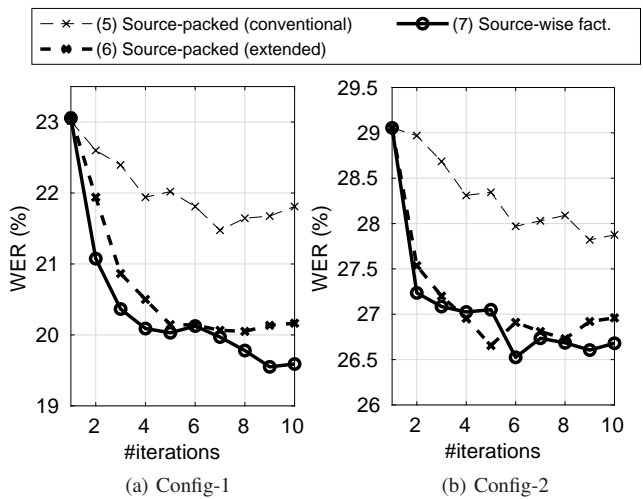}\\
{~\hspace{2mm}\footnotesize (a) Config-1\hspace{2.8cm}(b) Config-2}
\caption{WERs (\%) obtained  for REVERB-2MIX when jointly optimizing WPE+wMPDR based on {\MIMO} factorization (conventional/extended) and {\SW} factorization approaches.}\label{exp:joint}
\end{figure}

Figure~\ref{exp:joint} compares the WERs obtained using the three approaches with Config-1 and Config-2. Our proposed methods, i.e., (6) {\Mimo} factorization (extended) and (7) {\Sw} factorization, performed comparably well and both greatly outperformed (5) {\Mimo} factorization (conventional).  

\begin{table}[t]
\renewcommand{\arraystretch}{1.1}
\centering
\caption{Computing time required for processing a mixture utterance of length of 9.44 s in REVERB-2MIX. Computing time was measured by elapsed time on a Matlab interpreter.}\label{tbl:time}
\begin{tabular}{lcc}\hline
Method & \multicolumn{2}{c}{Time (s)}\\\cline{2-3}
& Config-1 & Config-2 
\\\hline
(4) {\Mimo} factorization (conventional) & 3467 
& 688\\
(5) {\Mimo} factorization (extended) & 209 
& 33\\
(6) {\Sw} factorization& 40 
& 23 \\\hline
\end{tabular}
\end{table}
Table~\ref{tbl:time} compares the computing times required for the three approaches to {estimate and apply the {\CBF}s with} ten estimation iterations for processing a mixture utterance whose length is 9.44~s. The computing time was measured by a Matlab interpreter as elapsed time.  {The computing times for estimating the masks were 0.63~s and 7.2~s with and without a GPU (NVIDIA 2080ti), and they are not included in the table.} As shown in the table, for both configurations, (6) {\Mimo} factorization (extended) greatly reduced the computing time in comparison with (5) {\Mimo} factorization (conventional), and (7) {\Sw} factorization further reduced the computing time.

The above results clearly demonstrate the superiority of the two proposed approaches over the conventional joint optimization technique in terms of both computational efficiency and estimation accuracy.  However, Table~\ref{tbl:time} indicates that the proposed approaches still require relatively large computing cost, {e.g., 40~s computing time for processing a 9.44~s utterance with Config-1,} to obtain the high performance gain shown in Fig.~\ref{exp:joint}~(a). Future work must address this problem. For example, it might be mitigated by setting the goal as extraction of a single target source. Then, due to the characteristics of {\SW} factorization, we can omit the estimation of the other sources, and omit the iterative estimation, e.g., when we separately estimate source variances using a neural network. As a reference, the computing time ($40$~s) in Table III required for the {\SW} factorization with Config-1 is roughly reduced to $2.0~\text{s}$ for one iteration per source (namely $40~\text{s}/10/2$), which results in the real-time factor being $0.21$ ($=2.0~\text{s}/9.44~\text{s}$).

\subsection{{Experiment-3: Evaluation using oracle masks}}\label{sec:exp3}

\begin{table}[!t]
\renewcommand{\arraystretch}{1.1}
\caption{{WER (\%) for RealData and CD (dB), FWSSNR (dB), PESQ, and STOI for SimData in REVERB-2MIX of enhanced signals obtained based on oracle masks using different beamformers after three estimation iterations with Config-1. Scores for REVERB-2MIX with no enhancement (No Enh) and those obtained by applying a wMPDR CBF, WPD \cite{MLWPD}, to REVERB (i.e., single speaker), are also shown.}}\label{tbl:om1}
\centering
\begin{tabular}{lm{6mm}m{4.5mm}m{6mm}m{6mm}m{5.5mm}}
\toprule
\multicolumn{1}{c}{Enhancement method}  & {WER} & \mcc{CD} & 
\multicolumn{1}{p{7mm}}{~\hspace{-2mm}FWSSNR} & {PESQ} & {STOI}\\
\midrule
No Enh (REVERB-2MIX) & 62.49 & 5.44 & ~~1.12 & ~1.12 & ~0.55\\
WPD (REVERB) \cite{MLWPD} & ~8.91 & 2.59 & ~~8.29 & ~2.41 & ~0.91 \\
\midrule
MPDR (w/o iteration) & 20.16 & 3.53 & ~~5.49 & ~1.86 & ~0.84 \\
MVDR (w/o iteration) & 20.32 & 3.56 & ~~5.36 & ~1.84 & ~0.83 \\
wMPDR & 20.12 & 3.31 & ~~6.11 & ~1.96 & ~0.86\\
\midrule
(1) WPE+MPDR (separate) & 12.89 & 3.39 & ~~6.11 & ~2.10 & ~0.87\\
(2) WPE+MVDR (separate) & 12.91 & 3.32 & ~~6.30 & ~2.07 & ~0.87\\
(3) WPE+wMPDR (separate) & 12.59 & 3.12 & ~~6.84 & ~2.21 & ~0.89 \\
(6) Source-packed fact. & {\bf 12.23} & 3.02 & ~~7.15 & {\bf ~2.33} & {\bf ~0.90} \\
(7) Source-wise fact. & {\bf 12.23} & {\bf 2.98} & {\bf ~~7.25} & {~2.32} & {\bf ~0.90} \\
\bottomrule
\end{tabular}
\end{table}

\begin{figure}[t]
\centering
\includegraphics[width=0.96\columnwidth]{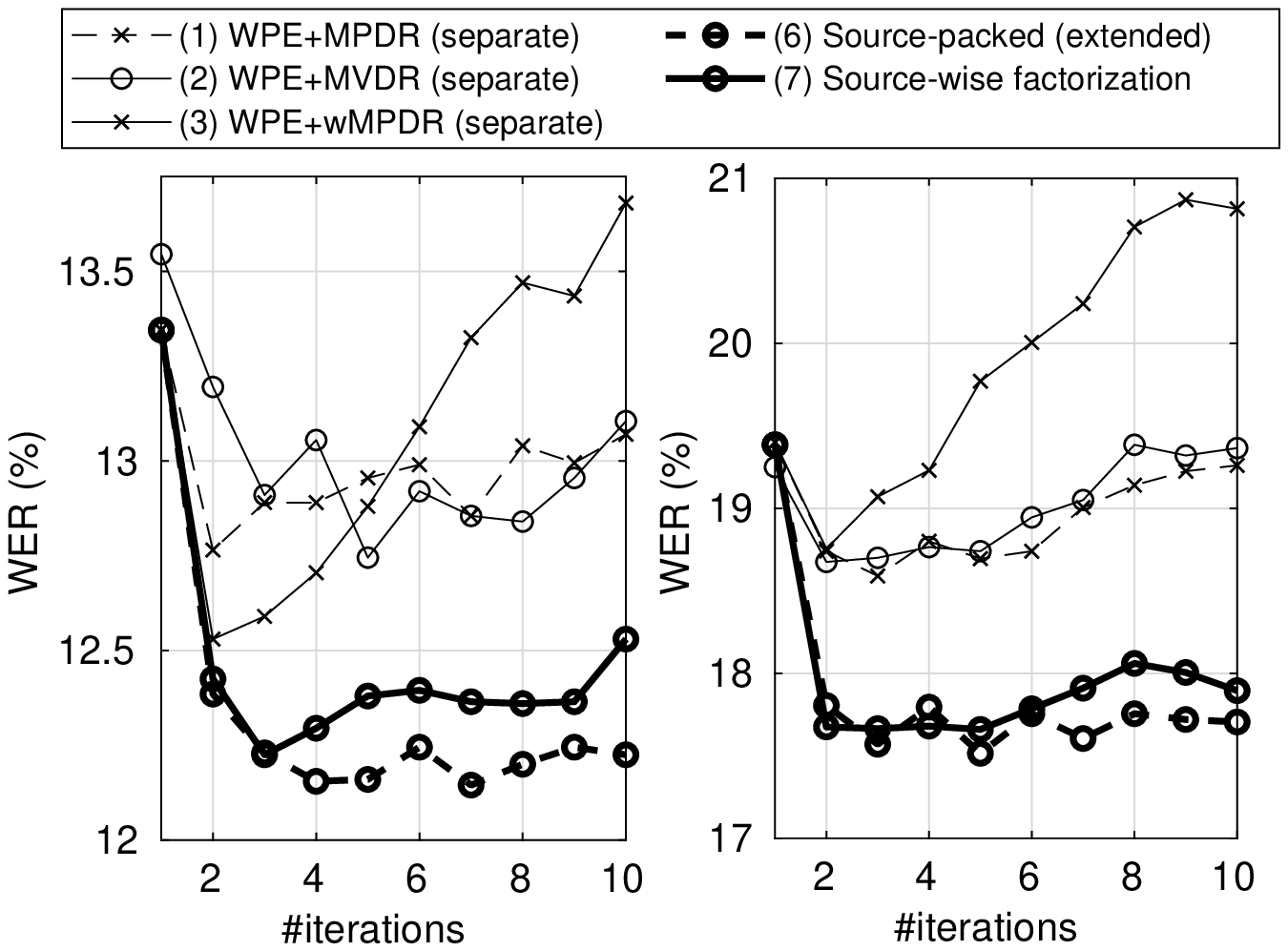}\\
{~\hspace{6mm}\footnotesize (a) REVERB-2MIX\hspace{2cm}(b) REVERB-3MIX}
\caption{{Comparison of WERs among cascade configuration and joint optimization approaches using Config-1 for REVERB-2MIX and REVERB-3MIX.}} \label{exp:wer3mix}
\vspace{5mm}
\includegraphics[width=0.96\columnwidth]{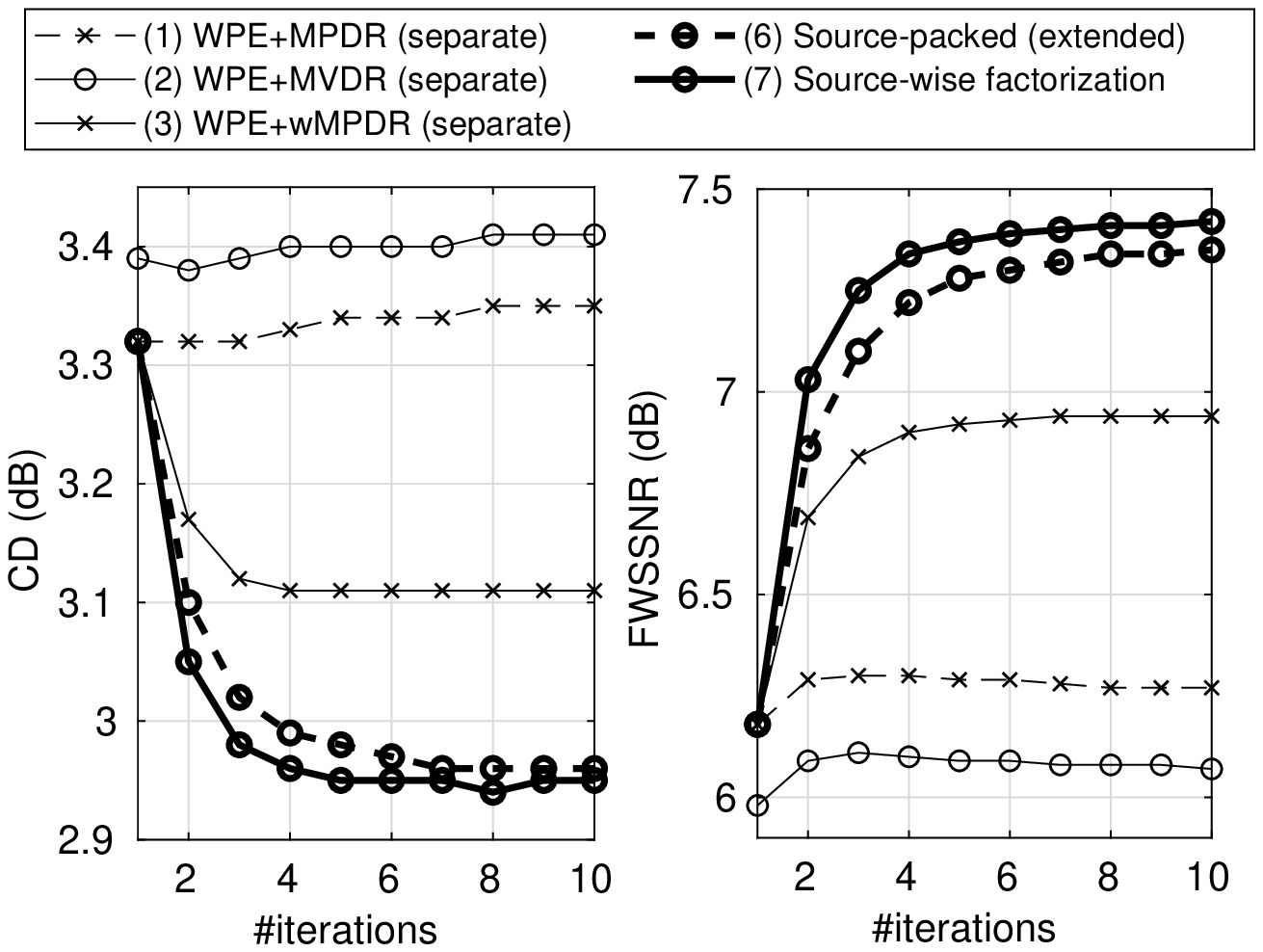}\\
~\hspace{-5mm}\includegraphics[width=0.96\columnwidth]{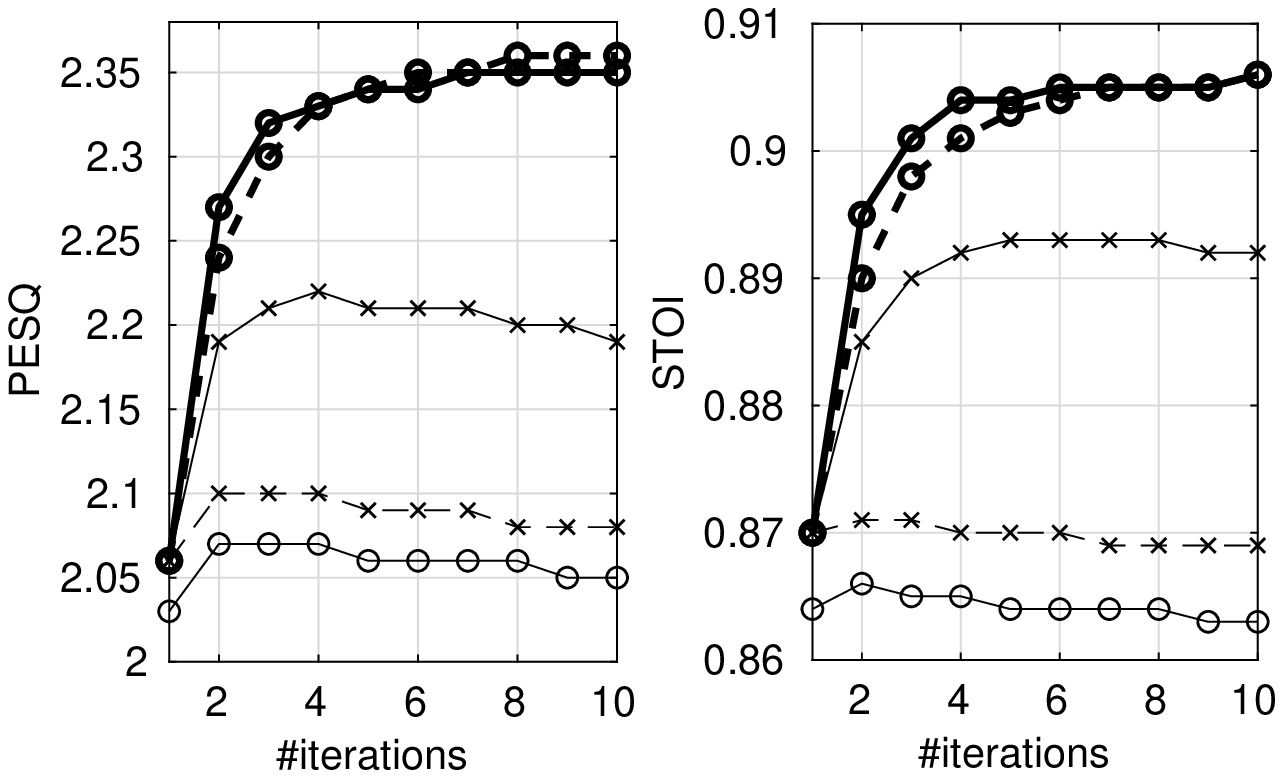}
\caption{{Comparison of CDs, FWSSNRs, PESQs, and STOIS among cascade configuration and joint optimization approaches using Config-1 for REVERB-2MIX.}} \label{exp:sepvsjoint3}
\end{figure}

{
In this experiment, we examined the performance of {\CBF}s using a different type of masks, i.e., oracle masks. An oracle mask, which is the power ratio of the desired signal to the observed signal at each TF point, is calculated using reference signals. Oracle masks can be precisely calculated for SimData in REVERB-2MIX using signal components in the observed signals. In contrast, we can only calculate the oracle masks approximately for RealData because we cannot access the signal components. Thus, we first estimated the desired signals by applying dereverberation and denoising to utterances in REVERB, and then calculated the oracle masks using the estimated desired signals for REVERB-2MIX and REVERB-3MIX.}

{Table~\ref{tbl:om1} shows WERs, CDs, FWSSNRs, PESQs, and STOIs measured on enhanced signals obtained from REVERB-2MIX using various (non-convolutional) beamformers and {\CBF}s after three estimation iterations. 
As a reference, the table also includes previously reported scores denoted by WPD (REVERB) \cite{MLWPD}, which were obtained by applying a wMPDR {\CBF}, referred to as WPD (see also Section~\ref{sec:swcbf} in this paper), to REVERB, i.e., noisy reverberant single speaker utterances.
In addition, the convergence curves obtained using the {\CBF}s in terms of WERs for REVERB-2MIX and REVERB-3MIX, and those obtained in terms of CDs, FWSSNRs, PESQs, and STOIs for REVERB-2MIX are respectively shown in Figs.~\ref{exp:wer3mix} and \ref{exp:sepvsjoint3}. In all these results, the two joint optimization approaches, (6) source-packed factorization (extended) and (7) source-wise factorization, outperformed all the other methods in terms of every measurement. As a whole, almost the same tendency was observed in the cases using the estimated masks. One exception is that the WERs obtained with the source-wise factorization tended to increase after a few iterations although such a tendency was not observed in terms of signal distortion measures. This means that improvement in the signal level distortion does not necessarily result in improvement in WER, and suggests the importance of optimization by ASR level criteria, similar to conventional beamforming techniques \cite{Jahn2017icassp,Ashwin2019waspaa}.
}

\section{Concluding remarks}\label{sec:conclusion}
This paper presented methods for optimizing a {\CBF} that  performs {\DNDRSS} based on ML estimation. We introduced two different approaches for factorizing a {\CBF}, i.e., {\MIMO} and {\SW} factorization approaches, and derived optimization algorithms for the respective approaches. A {\CBF} can be factorized without loss of optimality into a multiple-target WPE filter followed by wMPDR beamformers using the {\MIMO} factorization approach, and into a set of {\target} WPE filters followed by wMPDR beamformers using the {\SW} factorization approach. This paper also presented the overall processing flows for both approaches based on an assumption that TF masks are provided as auxiliary inputs. In the flows, the time varying source variances, which are required for ML estimation, can be optimally estimated jointly with the {\CBF} using iterative optimization; the steering vectors of the desired signals, which are required for beamformer optimization, can be reliably estimated based on the dereverberated multichannel signals obtained at an optimization step. 

Experiments using noisy reverberant sound mixtures show that the proposed optimization approaches substantially improved the {\CBF} performance in comparison with the conventional cascade configuration in terms of ASR performance and signal distortion reduction. Our proposed approaches can also greatly reduce the computing cost with improved estimation accuracy in comparison with the conventional joint optimization technique.
The proposed approaches, however, still result in relatively large computing costs to obtain high performance gain. Future work will address this problem.

\appendices


\section{Derivation of Eqs.~(\ref{eq:Psifast}) and (\ref{eq:psifast})}\label{sec:Psifast}
We can rewrite $\vect\Psi$ in Eq.~(\ref{eq:MS_wpe2}) using Eq.~(\ref{eq:Rq}):
\begin{align}
    \vect{\Psi}&=\frac{1}{T}\sum_t\OL{\vect{X}}_t^{\HT}\vect{\Phi}_{\vect{q},t}\OL{\vect{X}}_t,\\
    &=\frac{1}{T}\sum_t\sum_i\frac{1}{\lambda_t^{(i)}}\left(\left(\vect{q}^{(i)}\right)^{\HT}\OL{\vect{X}}_t\right)^{\HT}\left(\left(\vect{q}^{(i)}\right)^{\HT}\OL{\vect{X}}_t\right).\label{eq:appb1}
\end{align}
Using Eq.~(\ref{eq:Y1}), $\left(\vect{q}^{(i)}\right)^{\HT}\OL{\vect{X}}_t$ can further be rewritten:
\begin{align}
    \left(\vect{q}^{(i)}\right)^{\HT}\OL{\vect{X}}_t&=
    \left(\vect{q}^{(i)}\right)^{\HT}\left(\vect{I}_M\otimes\OL{\vect{x}}_t^{T}\right),\\
    &=\left(\vect{q}^{(i)}\right)^{\HT}\otimes \OL{\vect{x}}_t^{T}.
\end{align}
Substituting the above equation in Eq.~(\ref{eq:appb1}) yields
\begin{align}
    \vect{\Psi}&=\frac{1}{T}\sum_t\sum_i\frac{1}{\lambda_t^{(i)}}\left(\vect{q}^{(i)}\otimes\left(\OL{\vect{x}}_t^{\HT}\right)^{\top}\right)\left(\left(\vect{q}^{(i)}\right)^{\HT}\otimes \OL{\vect{x}}_t^{T}\right),\\
    &=\frac{1}{T}\sum_t\sum_i\frac{1}{\lambda_t^{(i)}}\left(\vect{q}^{(i)}\left(\vect{q}^{(i)}\right)^{\HT}\right)\otimes\left(\OL{\vect{x}}_t\OL{\vect{x}}_t^{\HT}\right)^{\top},\\
    &=\sum_i\left(\vect{q}^{(i)}\left(\vect{q}^{(i)}\right)^{\HT}\right)\otimes\left(\OL{\vect{R}}_{\vect{x}}^{(i)}\right)^{\top}.
\end{align}
Similarly, we can obtain
\begin{align}
    {\boldsymbol\psi}&=\frac{1}{\top}\sum_t\OL{\vect{X}}_t^{\HT}\vect{\Phi}_{\vect{q}}{\vect{x}}_t,\\
    &=\frac{1}{T}\sum_t\sum_i\frac{1}{\lambda_t^{(i)}}\left(\vect{q}^{(i)}\otimes \left(\OL{\vect{x}}_t^{\HT}\right)^{\top}\right)\left(\left(\vect{q}^{(i)}\right)^{\HT}\vect{x}_t\right),\\
    &=\frac{1}{T}\sum_t\sum_i\frac{1}{\lambda_t^{(i)}}\left(\vect{q}^{(i)}\otimes \left(\left(\vect{x}_t\OL{\vect{x}}_t^{\HT}\right)^{T}\left(\vect{q}^{(i)}\right)^{*}\right)\right),\\
    &=\sum_i\left(\vect{q}^{(i)}\otimes \left(\vect{P}_{\vect{x}}^{(i)}\vect{q}^{(i)}\right)^{*}\right).
\end{align}

\bibliographystyle{IEEEtran}
\bibliography{IEEEabrv,bibs}

\end{document}